\documentclass[12pt]{iopart}
\usepackage{epsfig}
\begin{document}

\topical[Application of the density matrix renormalization group method]
{Application of the density matrix renormalization group method to
finite temperatures and two-dimensional systems}
\author{Naokazu Shibata}
\address{Department of Basic Science, University of Tokyo, Komaba, 
Tokyo 153-8902 Japan}

\begin{abstract}
%\abst{
The density matrix renormalization group (DMRG) method 
and its applications to finite temperatures and two-dimensional 
systems are reviewed.
The basic idea of the original DMRG method, which allows
precise study of the ground state properties and 
low-energy excitations, is presented for models
which include long-range interactions.
The DMRG scheme is then applied to the diagonalization of the
quantum transfer matrix for one-dimensional systems, and a 
reliable algorithm at finite temperatures is formulated.
Dynamic correlation functions at finite temperatures 
are calculated from the 
eigenvectors of the quantum transfer matrix with analytical 
continuation to the real frequency axis.
An application of the DMRG method to two-dimensional 
quantum systems in a magnetic field is demonstrated and reliable
results for quantum Hall systems are presented.
%}
\end{abstract}
\maketitle

\section{Introduction}
Numerical calculations are now commonly used in various studies
in physics. 
The recent remarkable advance in computer technology and the
development of efficient algorithms provide us with a growing 
opportunity to resolve behaviors of various interacting systems
leading us to a new understanding of natural phenomena.
In condensed matter physics, one of the recent progresses has been
made by the density matrix renormalization group (DMRG) method, 
which was developed by Steven White in 1992 \cite{White,White2}.
This method is a type of variational method combined with a real space 
renormalization group method, which enables us to obtain the
ground-state wavefunction of large-size systems with controlled 
high accuracy. 
The DMRG method has excellent features
compared with other standard numerical methods.
In contrast to the quantum Monte Carlo method, 
the DMRG method is free from statistical errors and 
the negative sign problem which hampers convergence 
of physical quantities at low temperatures.
Furthermore, the essentially exact ground state of large systems  
extending the limitation of the exact diagonalization method
is obtained within a restricted number of basis states.
The truncation error caused by the restriction 
of basis states is systematically controlled by the density 
matrix calculated from the ground-state wavefunction and the 
obtained results are easily improved by increasing the number of 
basis states retained in the system.

The ground-state properties of frustrated quantum 
spin systems and the strongly correlated electron 
systems in one dimension
have been extensively studied by the DMRG method. 
Our understanding of these systems was greatly advanced and 
such successful calculations on one-dimensional quantum systems
have promoted many applications to other problems.
In 1995, Nishino applied the DMRG method to the transfer matrix
and he calculated the partition function of two-dimensional classical 
systems at finite temperatures \cite{Nishino}.
In 1996, Bursill {\it et al} 
applied this method to the quantum transfer matrix (QTM) for
$S=1/2$ XY spin chains,
and calculated thermodynamic quantities at finite temperatures.
%Their calculations on the non-Hermitian density 
%matrix was however not correct and reliable results 
%were obtained only down to $T/J=0.2$ \cite{Bursill}.
However, they used symmetric projection $|V^R\rangle \langle V^R|$ 
for the calculation of the density matrix and reliable results 
were obtained only down to $T/J=0.2$ \cite{Bursill}.
Correct calculations were done for $S=1/2$ Heisenberg 
spin chains by Wang and Xiang \cite{Wang}, and Shibata \cite{Shibata} in 1997,
and they obtained accurate results down to $T/J=0.01$.
The thermodynamic quantities and spin correlation length of 
the Heisenberg chains with $S=1/2,1$, and $3/2$ 
were systematically calculated by Xiang \cite{FT_spin}.
The stability of this method for fermion systems was
shown in the calculation of thermodynamic quantities of 
the one-dimensional Kondo lattice model,
which is a canonical model of heavy fermion systems \cite{FT_KLM}.
In 1999 the reliability of the method was also shown for 
frustrated quantum spin chains, $t$-$J$ ladders 
and the Kondo lattice model away from 
half-filling \cite{FT_Zigzag,FT_tJ_Lad,FT_KLM_d},
where the negative sign problem arises in the quantum Monte Carlo 
simulations.
Dynamic correlation functions and one-particle Green's functions 
in the Kondo lattice model were calculated at finite temperatures, 
and the temperature-induced
gap formation in the Kondo insulators was clarified \cite{Mutou}. 
There has been a review on the DMRG calculations of thermodynamic and 
dynamic quantities of the Kondo lattice model \cite{FT_KLM_rev}.
Autocorrelations in quantum spin chains were systematically 
studied by Naef {\it et al} and its reliability
was discussed \cite{Naef}.
The thermodynamic quantities of a spin-Peierls system of 
CuGeO$_3$ was studied by Kl\"umper {\it et al} \cite{FT_Spin_Pei}.
% \"
Rommer and Eggert applied the finite-temperature DMRG method to
impurity Kondo problems, and their results were shown to be
consistent with the predictions of field theory 
calculations \cite{FT_Imp_K}. The quasi-particle density of 
states and the dynamic spin and charge correlation
functions of the doped one-dimensional Kondo insulators were
calculated and their doping dependence was clarified \cite{Shibata_KLd}.
In 2000, Naef and Wang \cite{Naef2} 
studied the nuclear spin relaxation rate $1/T_1$ in 
the two-leg spin ladder and compared the results to theoretical 
predictions and experimental measures. 
Maeshima and Okunishi \cite{FT_zigzag_mag} calculated
thermodynamic quantities of frustrated quantum spin chains
under a magnetic field and studied behaviors near the critical fields 
in the ground-state magnetization process.
The magnetization process of the Heisenberg spin ladder 
was analyzed by Wang and Yu \cite{FT_SpinLAD_mag}, 
and they determined the phase diagram consisting
of disordered spin liquid, the Luttinger liquid, spin-polarized phases, 
and a classical regime.
Ammon and Imada \cite{ammon1,ammon2,ammon3} systematically studied
the low-temperature properties of doped $S=1$ spin chains
and clarified behaviors of various 
correlation functions.
In 2001, the thermodynamic properties of the $S=1/2$ Heisenberg chain 
with a staggered Dzyaloshinsky-Moriya interaction were studied and
anomalous behaviors in Yb$_4$As$_3$ were explained \cite{Yb4_theo,Yb4_exp}.
The thermodynamic properties of the $t$-$J$ chain
were studied by Sirker and Kl\"umper,
%\" 
and the crossover behavior to Tomonaga-Luttinger liquid was 
shown \cite{Sirk3,Sirk1}.
Maruyama {\it et al} \cite{Maru} studied
properties of a non-magnetic impurity in Kondo insulators.
Recently, Sirker and Khaliullin \cite{Sirk2} studied dimerization in 
a one-dimensional spin-orbital model with spins $S = 1$.

The application of the DMRG method to two-dimensional quantum 
systems is currently the most challenging subject and many 
algorithms have been proposed. 
Most of them use mappings on to effective
one-dimensional models with long-range interactions, and 
the standard DMRG method is applied to the effective 
one-dimensional systems.
In 1996, White \cite{White3} applied the DMRG method to a two-dimensional 
frustrated quantum spin model for CaV$_4$O$_9$
and showed the existence of the spin gap. 
In 1998, White and Scalapino \cite{White4} studied the two-dimensional 
$t$-$J$ model with a hole doping and found a striped phase.
They also studied the competition between stripes and 
pairing in an $n$-leg $t$-$t'$-$J$ model \cite{White5},
and made a critical analysis on 
the phase separation and stripe formation in the 
two-dimensional $t$-$J$ model \cite{White6}.
In 2001, Shibata and Yoshioka \cite{Shibata2,Shibata4} 
applied the DMRG method to 
quantum Hall systems and the ground state phase diagram 
of two-dimensional electrons in a high Landau level was determined.
Xiang {\it et al} \cite{Xiang} proposed an efficient mapping on to a
one-dimensional system which retains the topological characteristics of 
two-dimensional lattices. They applied this method to the $S=1/2$ 
Heisenberg model on both square and triangular lattices.
In 2003, the ground state phase diagram 
of two-dimensional electrons in the lowest and the second lowest
Landau levels were determined and the existence of 
various quantum liquids and charge ordered states, 
such as Laughlin state and the Wigner crystal,
was confirmed with new stripe states \cite{Shibata3}.

Another approach to higher dimensions was proposed by
Xiang in 1996 in which the Hamiltonian is renormalized
in momentum space \cite{Xiang2,Nishimoto}. 
This method is clearly efficient for weakly interacting systems. 
The DMRG method has also been applied to classical systems 
in higher dimensions \cite{Nishino}. 
In 1996 the DMRG method was
used to renormalize the corner transfer matrix and 
the corner transfer matrix renormalization group (CTMRG) method 
was formulated \cite{CTM1}.
This method was extended to three-dimensional classical 
models in 1998 \cite{CTM3}.
Recently the CTMRG method was extended to study one-dimensional
stochastic models \cite{Kem}. 
The DMRG algorithm has also been used to optimize tensor 
product states in three-dimensional classical systems, 
and this method was applied to two-dimensional quantum spin 
systems at finite temperatures \cite{Maeshima}.
There has been an analytical study on the spectra of the 
density matrices used in the DMRG calculations for two-dimensional 
systems \cite{Chun}.
Several topics of the DMRG calculations are summarized in
a book published in 1998 \cite{DMRG_book}.

In this topical review, the applications of the DMRG method
to finite temperatures and two-dimensional quantum systems are 
reviewed with various techniques in the calculation.
We first briefly summarize the basic idea of the 
DMRG method and explain how we treat long range interactions.
We then apply the DMRG scheme to the QTM and 
formulate an algorithm for the calculations of thermodynamic 
quantities and dynamic correlation functions at finite 
temperatures.
We also apply the DMRG algorithm to 
two-dimensional quantum systems in a magnetic field 
and calculate various ground states realized in 
quantum Hall systems.

\begin{figure}
\epsfxsize=75mm \epsffile{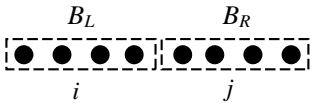}
\caption{\label{Fig_BL}
The system divided into two blocks $B_L$ and $B_R$, 
whose basis states are represented by indices $i$ and $j$. 
The solid circles represent local orbitals or quantum spins.
}
\end{figure}

\section{Zero temperature DMRG}
A numerical study of quantum many-body problems
requires the handling of exponentially increasing basis states of 
the system. For example, when we represent the 
Hamiltonian of two-spin system with $S=1/2$, four basis states,
$|\uparrow,\uparrow>$, $|\uparrow,\downarrow>$,
$|\downarrow,\uparrow>$, and $|\downarrow,\downarrow>$
are needed.
As is easily expected, the dimension of the Hamiltonian 
exponentially increases as $2^N$ with the increase of the number 
of spins $N$. This exponential dependence prevents exact 
numerical diagonalizations for more than 30 spins in usual cases. 
The DMRG method is designed to diagonalize the Hamiltonian
of large-size systems extending this limitation of 
the exact diagonalization by restricting the basis states.
Here we briefly review how the DMRG method
restricts the basis states with controlled accuracy.

\subsection{Restriction of basis states}
Let us consider a system consisting of two blocks, 
$B_L$ and $B_R$ (see figure \ref{Fig_BL}),
and represent the ground-state wavefunction $|\Psi \rangle$ 
using the basis states of the blocks described by 
$|i\rangle$ and $|j\rangle$: 
\begin{equation}
|\Psi \rangle = \sum_{i,j} \Psi_{ij} |i\rangle |j\rangle .
\end{equation}
Here we define the following density matrix $\rho$ 
for the block $B_L$, whose basis states are 
represented by $|i\rangle$:
\begin{equation}
\label{DME}
\rho^L_{i i'}=\sum_j  \Psi^*_{ij} \Psi_{i'j} .
\end{equation}
The norm of the wavefunction is then written by
\begin{equation}
\langle \Psi|\Psi \rangle = \sum_{i,j} \Psi^*_{ij} \Psi_{ij} = 
\sum_i \rho^L_{i i} =\mbox{Tr}\ \rho^L .
\end{equation}
The question is how we reduce the number of basis states
in the block $B_{L}$ while keeping $\mbox{Tr}\ \rho^L$ as much as
possible. 
Since $\mbox{Tr}\ \rho^L$ is equivalent to the sum of the
eigenvalues of $\rho^L$, we first solve the eigenvalue equation
\begin{equation}
w^\alpha v^\alpha_i = \sum_{i'} \rho^L_{i i'} v^\alpha_{i'}
\end{equation}
and transform the basis states 
from $|i\rangle$ to $|\alpha\rangle$ which is defined by
\begin{equation}
|\alpha\rangle = \sum_i v^\alpha_i |i\rangle .
\end{equation}
Since $\rho^L$ is Hermitian, the basis states  $|\alpha\rangle$ 
satisfy the orthogonality 
\begin{equation}
\langle \alpha |\alpha' \rangle =
\sum_i (v^\alpha_i)^* v^{\alpha'}_i = \delta_{\alpha \alpha'}.
\end{equation}

\begin{figure}
\epsfxsize=75mm \epsffile{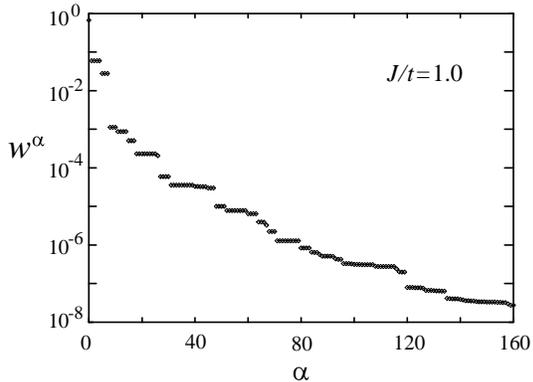}
\caption{\label{Fig_DM}
Eigenvalues of the density matrix obtained in 
the one-dimensional Kondo lattice model. 
The number of conduction electrons and localized spins
are both 32. The electrons and the
spins are coupled through on-site antiferromagnetic exchange 
coupling $J$. $t$ is the hopping integral.
}
\end{figure}

After this transformation, the density matrix becomes
diagonal and the norm of the wavefunction is written by
\begin{equation}
\langle \Psi|\Psi \rangle = \mbox{Tr}\ \rho^L
= \sum_\alpha w^\alpha.
\end{equation}
This result clearly shows that the norm is efficiently
preserved by keeping only eigenvectors whose eigenvalues $w^\alpha$ 
are large.
Thus the optimal $m$
basis states are obtained from the eigenstates of the $m$ largest 
eigenvalues of the density matrix.
Since $\rho^L$ is the product of
$\Psi_{ij}$ and its Hermitian conjugate $\Psi^*_{ij}$, 
the eigenvalues $w^\alpha$ are all positive and
we can sort $w^\alpha$ as
\begin{equation}
w^1 \ge w^2 \ge w^3 \ge ... \ge w^N \ge 0 
\end{equation}
where $N$ is the number of basis states in $B_L$.
The truncation error, Er($m$), of the wavefunction 
represented by $m$ basis states in $B_L$ is then given by
\begin{equation}
\mbox{Er}(m) = \sum_{\alpha=m+1}^{N}w^\alpha = 
1- \sum_{\alpha=1}^{m}w^\alpha.
\end{equation}
Here we have used 
$\langle \Psi|\Psi \rangle = \sum_{\alpha=1}^{N} 
w^\alpha = 1$.
Thus the  truncation error depends on the distribution of 
the eigenvalues of the density matrix, and high accuracy is 
obtained 
when $w^\alpha$ decays rapidly with increasing $\alpha$.
A typical example of $w^\alpha$ obtained for 
the one-dimensional Kondo lattice model is shown in figure \ref{Fig_DM}.
Even though this model consists of conduction electrons and 
localized spins, the accuracy of $10^{-7}$ is attained
by keeping only 150 states in each block. 
Accurate results in the DMRG calculations are obtained when
$w^\alpha$ decays almost exponentially as shown in figure \ref{Fig_DM}.

\begin{figure}
\epsfxsize=75mm \epsffile{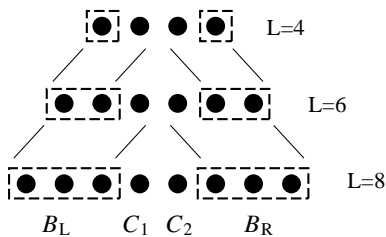}
\caption{\label{Fig_InfDMRG}
Schematic diagram for infinite system algorithm of the DMRG.
}
\end{figure}

\subsection{Infinite system algorithm}
Here we describe how we obtain the Hamiltonian of a large system
within a restricted number of basis states. 
In the infinite system algorithm of the DMRG method\cite{White},
we start from a small system and iteratively 
extend the system by adding new sites between the two blocks,
as schematically shown in figure \ref{Fig_InfDMRG}.
At each extension of the blocks we restrict the basis states
using the density matrix defined in section 2.1.
In the following, we describe the algorithm of the DMRG method
applied to one-dimensional quantum systems which have
long-range interactions.

We usually start the calculation from a system consisting of four 
single sites with open boundary conditions. 
The left and right blocks contain the single 
site at one end of the system, as shown in figure \ref{Fig_InfDMRG}. 
The Hamiltonian is written by
\begin{eqnarray}
\label{Ham}
H &=&H^{B_L}+H^{C_1} + H^{C_2}+ H^{B_R} \nonumber \\         
  & &  + H^{B_L-C_1} + H^{C_1-C_2}+ H^{C_2-B_R} \nonumber \\
  & &  + H^{B_L-C_2} + H^{C_1-B_R} + H^{B_L-B_R}
\end{eqnarray}
where $H^{B_L}$ and $H^{B_R}$ contains operators 
represented by only the basis states of the left and right  
blocks, respectively, and $H^{C_1}$ and $H^{C_2}$ are the
single-site Hamiltonian between the two blocks.
The interaction and hopping terms between different blocks and 
single sites are represented by  $H^{B_L-C_1} + H^{C_1-C_2} + H^{C_2-B_R}$ and 
$ H^{B_L-C_2} + H^{C_1-B_R} + H^{B_L-B_R}$, where the former three terms 
describe the nearest-neighbor interactions and hoppings,
and the latter three terms represent next-nearest-neighbor and further 
long-range interactions and hoppings. 
In order to avoid complexity we have assumed that each
inter-site interaction and hopping term is written by operators 
at two different sites, and there are no multi-site interactions
such as ${\bf S}_l \cdot ( {\bf S}_m \times {\bf S}_n)$ with $l\ne m \ne n$.
The Heisenberg model with nearest-neighbor interaction $J_1$
and next-nearest-neighbor interaction $J_2$
in magnetic field $h^z$ is written by 
$H^{B_L}= h^z S^z_1$, $H^{C_1} = h^z S^z_2$, $H^{C_2}= h^z S^z_3$, 
$H^{B_R}=h^z S^z_4$,
$H^{B_L-C_1} = J_1{\bf S}_1 \cdot {\bf S}_{2}$,
$H^{C_1-C_2} = J_1{\bf S}_2 \cdot {\bf S}_{3}$,
$H^{C_2-B_R} = J_1{\bf S}_3 \cdot {\bf S}_{4}$,
$ H^{B_L-C_2} =  J_2{\bf S}_1 \cdot {\bf S}_{3}$,
$H^{C_1-B_R} =  J_2{\bf S}_2 \cdot {\bf S}_{4}$,
and $H^{B_L-B_R}=0$.

We then calculate the ground state-wavefunction $|\Psi\rangle$
by diagonalizing the above Hamiltonian.
The obtained wavefunction is represented in the basis states of the 
blocks and single sites as
\begin{equation}
|\Psi \rangle = \sum_{i_L i_1 i_2 i_R} \Psi_{i_L i_1 i_2 i_R} 
|i_L\rangle |i_1\rangle |i_2\rangle |i_R\rangle 
\end{equation}
where $|i_L\rangle$ and $|i_R\rangle$ are the basis states 
of $B_L$ and $B_R$, and $|i_1\rangle$ and $|i_2\rangle$
are the basis states of $C_1$ and $C_2$.
We next extend the blocks by including the neighboring single site.
The basis states 
%$|i_L^n\rangle$ and $|i_R^n\rangle$ 
of the extended blocks $B_L\otimes C_1$ and $B_R\otimes C_2$
are represented by the product space of the original block and 
the single site,
$|i_L\rangle|i_1\rangle$ and $|i_R\rangle|i_2\rangle$.
We then restrict the basis states of the extended blocks.
%to avoid exponential increase of the basis states.
For this purpose we calculate the density matrices for 
the extended blocks, $\rho^{L}$ and $\rho^{R}$ defined in section 2.1
\begin{eqnarray}
\rho^{L}_{i_L i_1, i_L' i_1'}&=& \sum_{i_R i_2} \Psi_{i_L i_1 i_2 i_R} 
\Psi^*_{i_L' i_1' i_2 i_R} \\
\rho^{R}_{i_R i_2, i_R' i_2'}&=& \sum_{i_L i_1} \Psi_{i_L i_1 i_2 i_R} 
\Psi^*_{i_L i_1 i_2' i_R'} .
\end{eqnarray}
We diagonalize the density matrices to obtain the eigenvectors 
$\vec{v}^{\alpha}$ and the corresponding eigenvalues $w^{\alpha}$ 
which satisfy 
\begin{eqnarray} \label{eig1}
w^{\alpha_L}v^{\alpha_L}_{i_L i_1} &=& 
\sum_{i_L' i_1'} \rho^{L}_{i_L i_1, i_L' i_1'} v^{\alpha_L}_{i_L' i_1'} \\
w^{\alpha_R}v^{\alpha_R}_{i_R i_2} &=& 
\sum_{i_R' i_2'} \rho^{R}_{i_R i_2, i_R' i_2'} v^{\alpha_R}_{i_R' i_2'} .
\end{eqnarray}
The new basis states of the extended blocks are obtained from the 
eigenvectors of the $m$ largest eigenvalues $|\alpha_L\rangle$ 
and $|\alpha_R\rangle$ defined by
\begin{eqnarray}
|\alpha_L\rangle &=& \sum_{i_L i_1} v^{\alpha_L}_{i_L i_1} 
|i_L\rangle |i_1\rangle\\
|\alpha_R\rangle &=& \sum_{i_R i_2} v^{\alpha_R}_{i_R i_2} 
|i_R\rangle |i_2\rangle. 
\end{eqnarray}
We then represent all operators using these new basis states.
The spin operator at the first site $(S_1^z)_{i_L i_L'}$ 
in the left block $B_L$ is given by
\begin{equation}
(S_1^z)_{\alpha_L {\alpha_L}'} =  \sum_{i_L {i_L}' i_1} 
v^{\alpha_L}_{i_L i_1} (v^{{\alpha_L}'}_{{i_L}' {i_1}})^* 
(S_1^z)_{i_L {i_L}'}. \label{op1}
\end{equation}
Similarly, the spin operators at the second site $(S_2^z)_{i_1 i_1'}$
which has been included in the new block ($B_L$){\scriptsize new} 
is represented by
\begin{equation}
(S_2^z)_{\alpha_L {\alpha_L}'} =  \sum_{i_L i_1 {i_1}'} 
v^{\alpha_L}_{i_L i_1} (v^{{\alpha_L}'}_{{i_L} {i_1}'})^* 
(S_2^z)_{i_1 {i_1}'}. \label{op2}
\end{equation}
The block Hamiltonian $(H^{B_L})_{\mbox{\scriptsize new}}
=H^{B_L}+H^{C_1}+H^{B_L-C_1}$ and the interaction between the new block 
and the single site $(H^{B_L-C_1})_{\mbox{\scriptsize new}}=
H^{C_1-C_2}+H^{B_L-C_2}$
are obtained by
\begin{eqnarray}
(H^{B_L}_{\alpha_L \alpha_L'})_{\mbox{\scriptsize new}}&=&
\sum_{i_L {i_L}' i_1 {i_1}'}
v^{\alpha_L}_{i_L i_1} (v^{{\alpha_L}'}_{{i_L}' {i_1}'})^* 
\left\{ H^{B_L}_{i_L i_L'}\delta_{i_1 i_1'}
+H^{C_1}_{i_1 i_1'}\delta_{i_L i_L'}
+H^{B_L-C_1}_{i_L i_1 i_L' i_1'}\right\} \nonumber\\
&&\\
(H^{B_L-C_1}_{\alpha_L i_2 \alpha_L' i_2'})_{\mbox{\scriptsize new}}
&=&\sum_{i_L {i_L}' i_1 {i_1}'}
v^{\alpha_L}_{i_L i_1} (v^{{\alpha_L}'}_{{i_L}' {i_1}'})^* 
\left\{ H^{C_1-C_2}_{i_1 i_2 i_1' i_2'}\delta_{i_L i_L'}
+H^{B_L-C_2}_{i_L i_2 i_L' i_2'}\delta_{i_1 i_1'}\right\}.
\end{eqnarray}
Similarly, $(H^{B_R})_{\mbox{\scriptsize new}}$ and 
$(H^{C_2-B_R})_{\mbox{\scriptsize new}}$
are given by
\begin{eqnarray}
(H^{B_R}_{\alpha_R \alpha_R'})_{\mbox{\scriptsize new}}&=&
\sum_{i_R {i_R}' i_2 {i_2}'}
v^{\alpha_R}_{i_R i_2} (v^{{\alpha_R}'}_{{i_R}' {i_2}'})^* 
\left\{ H^{B_R}_{i_R i_R'}\delta_{i_2 i_2'}
+H^{C_2}_{i_2 i_2'}\delta_{i_R i_R'}
+H^{C_2-B_R}_{i_R i_2 i_R' i_2'}\right\} \nonumber\\ 
&&\\
(H^{C_2-B_R}_{\alpha_R i_1 \alpha_R' i_1'})_{\mbox{\scriptsize new}}
&=&\sum_{i_R {i_R}' i_2 {i_2}'}
v^{\alpha_R}_{i_R i_2} (v^{{\alpha_R}'}_{{i_R}' {i_2}'})^* 
\left\{ H^{C_1-C_2}_{i_1 i_2 i_1' i_2'}\delta_{i_R i_R'}
+H^{C_1-B_R}_{i_R i_1 i_R' i_1'}\delta_{i_2 i_2'}\right\}.
\end{eqnarray}
The long range interactions 
$(H^{B_L-C_2})_{\mbox{\scriptsize new}}$,
$(H^{C_1-B_R})_{\mbox{\scriptsize new}}$ 
and $(H^{B_L-B_R})_{\mbox{\scriptsize new}}$ are not simply 
obtained by transforming the previous Hamiltonian, 
because they are defined after we add new single sites
between the two blocks.
These terms are constructed from the operators represented by 
the basis states of the new blocks and the single sites added in the system.
The new Hamiltonian of six sites is then written by
\begin{eqnarray}
(H)_{new} &=& (H^{B_L})_{\mbox{\scriptsize new}}
  +H^{C_1} + H^{C_2} + (H^{B_R})_{\mbox{\scriptsize new}} \nonumber \\         
  & &  + (H^{L-C1})_{\mbox{\scriptsize new}}  
+ H^{C_1-C_2}+ (H^{C_2-B_R})_{\mbox{\scriptsize new}}  \nonumber \\
  & &  + (H^{B_L-C_2})_{\mbox{\scriptsize new}}  
+ (H^{C_1-B_R})_{\mbox{\scriptsize new}}  
+ (H^{B_L-B_R})_{\mbox{\scriptsize new}} .
\end{eqnarray}
This Hamiltonian has similar structure to the previous one written in 
equation (\ref{Ham}),
and we repeat the above procedure until we obtain the desired size of system.
Since there is no limitation on the size of system, this
algorithm is called the infinite system algorithm of the DMRG.

\begin{figure}
\epsfxsize=75mm \epsffile{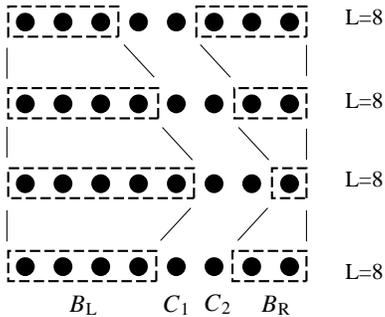}
\caption{\label{Fig_FinDMRG}
Schematic diagram for finite system algorithm of the DMRG.
}
\end{figure}

\subsection{Finite system algorithm}
In the infinite system algorithm of the DMRG we calculate the Hamiltonian 
of any size of systems by iteratively extending the blocks. 
However, in order to deal with exponential increase in the number the 
basis states, we truncate the basis states.
This truncation introduces an artificial error in the wavefunction,
and we need to minimize the error by reconstructing 
the basis states within a fixed size of system.
The algorithm to find the optimal basis states in a finite 
system is called the finite system algorithm of the DMRG.
Here, we describe how we improve the ground-state wavefunction
using the finite system algorithm of the DMRG.

Suppose we have extended the system up to $2N+2$ sites by
using infinite system algorithm of the DMRG. The left and 
right blocks now contain $N$ sites and
the system is written by
\begin{equation}
\mbox{B}_L(N) \bullet \bullet \ \mbox{B}_R(N).
\end{equation}
Here $B_L(N)$ and $B_R(N)$ represent the left and 
the right blocks which contain $N$ sites
and $\bullet$ represents a single site between the two blocks.
In order to reconstruct the right block first, we extend 
only the left block and reduce the right block, 
as shown in figure \ref{Fig_FinDMRG}.
We calculate the optimal basis states for $B_L(N+1)$ 
by diagonalizing the density matrix calculated from the ground-state 
wavefunction of the system $B_L(N) \bullet \bullet B_R(N)$,
and we obtain the Hamiltonian $H^{B_L}$ for the left block $B_L(N+1)$ 
by a similar calculation to that in the infinite system algorithm 
of the DMRG. We do not need to calculate the basis states and
the Hamiltonian $H^{B_R}$ for $B_R(N-1)$ because they have already been 
obtained in the infinite system algorithm 
of the DMRG. The system of $2N+2$ sites is now represented by
\[
\mbox{B}_L(N+1) \bullet \bullet  \ \mbox{B}_R(N-1).
\]
We repeat the extension of the left block until the right block 
becomes single site. Then, we extend the right block from the 
single site to reconstruct the basis states in $B_R$.
Now the basis states of the right block are always 
determined from the ground-state wavefunction of $2N+2$ sites. 
This is different from the calculation in the infinite system 
algorithm, where the size of system 
increases with the extension of the blocks. This difference
improves the ground-state wavefunction obtained
in the infinite system algorithm of the DMRG.

The extension of the right block is continued until the left block is 
reduced down to the single site. We then extend the left block 
to reconstruct the basis states in $B_L$.
We repeat such sweeps until the ground-state 
energy and its wavefunction converge.
Since we have reduced Hilbert space of the Hamiltonian
by restricting the basis states,
the obtained ground-state energy is always higher than the
exact one. The increase in the number of the basis states 
systematically lowers the ground-state energy and its value
approaches to the exact one.
Thus the finite system algorithm of the DMRG method
is equivalent to a variational method in which a trial wavefunction 
is constructed automatically under the restriction on the
number of basis states.

\subsection{Evaluation of physical quantities}
We evaluate various physical quantities from the wavefunction 
obtained by the DMRG calculations. 
Since the basis states of the wavefunction 
alter after the reconstructions of the blocks,
we need the operators represented in the basis states
of the wavefunction.
These operators are obtained by successive transformations 
of basis states
starting from the operators defined in the initial single site.
At each reconstruction of the blocks
the operators such as $S^z_i$ is transformed as follows.
If the $i$th site is the single site next to the block $B_L$,
$(S_i^z)_{i_1 {i_1}'}$ is transformed by
\begin{equation} 
\label{op3}
(S_i^z)_{\alpha_L {\alpha_L}'} =  \sum_{i_L i_1 {i_1}'} 
v^{\alpha_L}_{i_L i_1} (v^{{\alpha_L}'}_{{i_L} {i_1}'})^* 
(S_i^z)_{i_1 {i_1}'} 
\end{equation}
where extended new left block include the $i$th site and has
basis states $|\alpha_L \rangle$, and 
$v^{\alpha_L}_{i_L i_1}$ is the
eigenvectors of the density matrix defined in equation (\ref{eig1}). 
After the $i$th site is included in the block $B_L$, $S^z_i$ is
transformed by
\begin{equation}
(S_i^z)_{\alpha_L {\alpha_L}'} =  \sum_{i_L {i_L}' i_1} 
v^{\alpha_L}_{i_L i_1} (v^{{\alpha_L}'}_{{i_L}' {i_1}})^* 
(S_i^z)_{i_L {i_L}'}. \label{op4}
\end{equation}
The product of the two operators at sites $i$ and $j$,  such as $S_i^z
S_j^z$ are transformed by
\begin{equation}
(S_i^z S_j^z)_{\alpha_L {\alpha_L}'} =  \sum_{i_L {i_L}' i_1 {i_1}'} 
v^{\alpha_L}_{i_L i_1} (v^{{\alpha_L}'}_{{i_L}' {i_1}'})^* 
(S_i^z)_{i_L {i_L}'} (S_j^z)_{i_1 {i_1}'} \label{op5}
\end{equation}
where the $j$th site is the single site next to the original left block $B_L$
and the $i$th site is in the left block $B_L$.
If both the $i$th and $j$th sites are in the left block
$B_L$, they are transformed by   
\begin{equation}
(S_i^z S_j^z)_{\alpha_L {\alpha_L}'} =  \sum_{i_L {i_L}' i_1} 
v^{\alpha_L}_{i_L i_1} (v^{{\alpha_L}'}_{{i_L}' {i_1}})^* 
(S_i^z S_j^z)_{i_L {i_L}'}. \label{op6}
\end{equation}

After we have obtained the operators represented by the basis states of
the wavefunction $\Psi_{i_L i_1 i_2 i_R}|
i_L\rangle|i_1\rangle|i_2\rangle|i_R\rangle$,     
we evaluate physical quantities. 
The local quantities 
$\langle \Psi | S_i^z |\Psi\rangle$ and
correlation functions 
$\langle \Psi | S_i^z S_j^z |\Psi\rangle$
are given by
\begin{eqnarray}
\langle \Psi | S_i^z |\Psi\rangle &=& \sum_{i_L {i_L}' i_1 i_2 i_R}
{\Psi^*_{{i_L}' i_1 i_2 i_R}} (S_i^z)_{{i_L}' i_L} 
{\Psi_{{i_L} i_1 i_2 i_R}} \\
\langle \Psi | S_i^z S_j^z |\Psi\rangle &=& \sum_{i_L {i_L}' i_1 i_2 i_R}
{\Psi^*_{{i_L}' i_1 i_2 i_R}} (S_i^z S_j^z)_{{i_L}' i_L} 
{\Psi_{{i_L} i_1 i_2 i_R}}
\end{eqnarray}
where both $S_i$ and $S_j$ are assumed to be in the left block.
If $S_j$ is in the right block while $S_i$ is in the left block,
the correlation function
$\langle \Psi | S_i^z S_j^z |\Psi\rangle$ is given by
\begin{equation}
\langle \Psi | S_i^z S_j^z |\Psi\rangle = \sum_{i_L {i_L}' i_1 i_2 i_R {i_R}'}
{\Psi^*_{{i_L}' i_1 i_2 {i_R}'}} (S_i^z)_{{i_L}' i_L} 
(S_j^z)_{{i_R}' i_R} {\Psi_{{i_L} i_1 i_2 i_R}}.
\end{equation}

\begin{figure}
\epsfxsize=75mm \epsffile{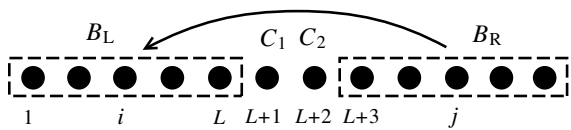}
\caption{\label{Fig_hop}
Long range hopping between the two blocks.
}
\end{figure}

\subsection{Fermion sign}
When an electron hops from one site to another,
the fermion sign appears in the matrix element
between the two basis states.
The fermion sign depends on the number of electrons between the 
two sites, and we need to count the electrons.
However, the number of electrons at each site
in the basis states of the blocks 
is neither one nor zero 
after we have transformed the basis states
using the eigenvectors of the density matrix.
In order to correctly deal with the fermion sign,
it is useful to define new creation 
(annihilation) operators, $\bar{c}_{i \sigma}^\dagger$ and
$(\bar{c}_{i \sigma})$,
which include the fermion sign generated by the hopping
from the $i$th site to one end of the block.
If we align the up-spin electrons and
down-spin electrons separately,  $\bar{c}_{i \sigma}^\dagger$
in the left block are defined by
\begin{equation}
\bar{c}_{i \sigma}^\dagger 
\equiv c_{i \sigma}^\dagger (-1)^{N_{i+1 \sigma}+N_{i+2 \sigma}+ 
\cdots +N_{L \sigma}}
\end{equation}
where $L$ is the number of sites in the left block 
(see figure \ref{Fig_hop}) and
$N_{i \sigma}$ are the number operators of the electron 
with spin $\sigma$ at the $i$th site.
The operators  $\bar{c}_{j \sigma}^\dagger$
in the right block are defined by 
%$\bar{c}_{j \sigma}^\dagger$ as
\begin{equation}
\bar{c}_{j \sigma}^\dagger \equiv c_{j \sigma}^\dagger 
(-1)^{N_{L+3 \sigma}+N_{L+4 \sigma}+ \cdots +N_{j-1 \sigma}}.
\end{equation}
The matrix element 
$\langle k| c^\dagger_{i \sigma} c_{j \sigma} | l \rangle$
generated by a long-range hopping between the two blocks is then given by
\begin{equation}
\langle k| c^\dagger_{i \sigma} c_{j \sigma} | l \rangle = (\bar{c}^\dagger_{i \sigma})_{k_L l_L} 
(\bar{c}_{j \sigma})_{k_R l_R} 
(-1)^{(N_{L+1 \sigma})_{l_{1}}+(N_{L+2 \sigma})_{l_{2}}} 
\delta_{k_{1} l_{1}} \delta_{k_{2} l_{2}}
\end{equation}
where $| l \rangle \equiv | l_L \rangle |l_{1} \rangle| l_{2} 
\rangle| l_R \rangle$. The coefficient
$(-1)^{(N_{L+1 \sigma})_{l_{1}}+(N_{L+2 \sigma})_{l_{2}}}$ 
is the fermion sign coming from the electrons with spin $\sigma$
in the two single sites between the blocks.
In this expression we need not calculate the fermion sign 
coming from the electrons in the blocks, and the 
matrix elements of long-range hoppings between the two 
blocks are obtained by only counting electrons in the 
two single sites between the two blocks.

The operators $\bar{c}_{i \sigma}^\dagger$
are obtained by successive transformations starting from the 
original operators defined in the single site.
When the left block is enlarged by including the $i_1$th site,
$\bar{c}^\dagger_{i \sigma}$ in the left block is transformed as
\begin{equation}
(\bar{c}^\dagger_{i \sigma})_{\alpha_L {\alpha_L}'} =  \sum_{i_L {i_L}' i_1} 
v^{\alpha_L}_{i_L i_1} (v^{{\alpha_L}'}_{{i_L}' {i_1}})^* 
(\bar{c}^\dagger_{i \sigma})_{i_L {i_L}'} (-1)^{N_{i_1 \sigma}} \label{opc}
\end{equation}
where $(-1)^{N_{i_1 \sigma}}$ is the fermion sign 
generated by the electron at the $i_1$th site,
and $\bar{c}^\dagger_{i \sigma} \equiv {c}^\dagger_{i \sigma}$
when the $i$th site is the right end of the left block, $i=L$.
Since the basis states $|l_{1}\rangle$ of the single site are the 
eigenstates of the number operator $N_{i_1}$, $N_{i_1}$ is either 
one or zero depending on the index $l_{1}$.

\subsection{Conserved quantities}
In the DMRG calculation, most of the computing time is 
consumed for the diagonalization of the Hamiltonian. 
Thus, it is important to use an efficient method for the diagonalization.
Here we consider the basis states classified by the
conserved quantities of the Hamiltonian. 
By using these basis states we need only diagonalize the 
Hamiltonian within a subspace specified by a set of quantum numbers.
For example, Heisenberg model without transverse magnetic
field conserves the $z$-component of the total spin, and all the
eigenstates of the Hamiltonian are classified by total $S^z$. 
The blocks $B_L(2)$ and $B_R(2)$ containing two $S=1/2$ spins
have basis states
$|\uparrow,\uparrow>$, $|\uparrow,\downarrow>$,
$|\downarrow,\uparrow>$, and $|\downarrow,\downarrow>$ 
with quantum numbers of $S^z=1,0,0$, and -1, respectively,
and the basis states of the total system 
$B^L(2) \bullet \bullet B^R(2)$
are constructed from various combinations of local basis states 
which satisfy 
\begin{equation}
S^z_{tot}=S^z_{L}+S^z_{C1}+S^z_{C2}+S^z_{R}=
\mbox{const.}
\end{equation}
When we calculate the ground state of $S^z_{tot}=0$,
there are 10-configurations which are
represented by ($S^z_{L},S^z_{C1},S^z_{C2},S^z_{R}$) =
(1,1/2,-1/2,-1),  (1,-1/2,1/2,-1), 
(1,-1/2,-1/2,0), (0,-1/2,-1/2,1), (0,1/2,-1/2,0),
(0,-1/2,1/2,0), (-1,1/2,1/2,0), (0,1/2,1/2,-1), 
(-1,1/2,-1/2,1), (-1,-1/2,1/2,1).
The number of total basis states is 20, which
is lower than $2^6=64$ of full Hilbert space.
Since the Hamiltonian $H^{B_L}$ and $H^{B_R}$ conserve 
total $S^z$, they are also block diagonal, and
$H^{B_L}+H^{C_1}+H^{C_2}+H^{B_R}$ have matrix elements only within 
each subspace specified by ($S^z_{B_L},S^z_{C_1},S^z_{C_2},S^z_{B_R}$).
The matrix elements between different subspaces are
generated only by the terms $H^{B_-C_1} + H^{C_1-C_2} + H^{C_2-B_R} 
+ H^{B_L-C_2} + H^{C_1-B_R} + H^{B_L-B_R}$.

The density matrix 
%\begin{equation}
$\rho_{i i'}=\sum_j \Psi^*_{i j} \Psi_{i' j} $
%\end{equation}
obtained from the ground-state wavefunction
is also block diagonal, because total $S^z$ of the 
basis states $j$ is uniquely determined by the basis states
$i$ of the density matrix. This means that  
the density matrix does not have any matrix elements
between different subspaces of total $S^z$.
Since the density matrix is block diagonal, 
the unitary transformation of the basis states
is defined only within the same subspace of total $S^z$, 
and the basis states of the new blocks are also classified 
by total $S^z$. 

By using the basis states classified by the conserved quantities
such as the z-component of the total spin and the number of electrons,
the dimension of the Hamiltonian, operators, and density matrix 
are reduced. The reduction of the dimensions improves the 
efficiency and accuracy of the numerical calculation, 
and reduces the memory space needed in the calculation.

\begin{figure}
\epsfxsize=75mm \epsffile{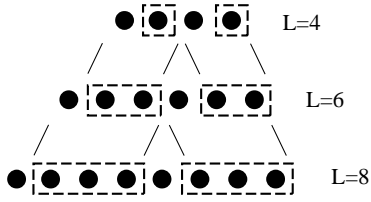}
\caption{\label{Fig_PBC}
Arrangement of the blocks and the single sites 
for the periodic boundary conditions.
}
\end{figure}

\subsection{Periodic boundary conditions}
\begin{figure}
\epsfxsize=75mm \epsffile{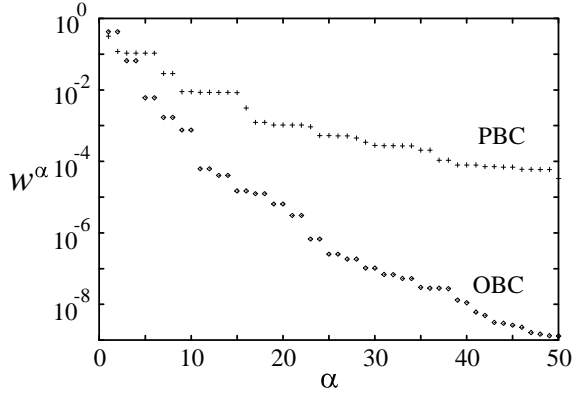}
\caption{\label{Fig_BC}
Eigenvalues of the density matrix for S=1/2 Heisenberg model
with 40 spins under the periodic boundary conditions (PBC)
and the open boundary conditions (OBC).
}
\end{figure}

In the DMRG calculations, the open boundary conditions are usually used.
%to obtain accurate results. 
However, the existence of the 
boundaries sometimes causes large boundary effects on the 
ground state. These boundary effects are removed when
we impose the periodic boundary conditions. 
In this case, we usually align the blocks and single sites symmetrically
as shown in figure \ref{Fig_PBC}. 
The truncation error under the periodic boundary conditions 
is generally large
compared with that obtained in the open boundary conditions.
To keep the accuracy of the results,  
we need much more basis states in each block.
This is actually shown in figure \ref{Fig_BC}, which shows 
slower decay of the eigenvalues of the density matrix for
the periodic boundary conditions.

Under the periodic boundary conditions, the left and 
right blocks are 
connected with each other through both ends of the blocks.
The reduction of the accuracy is caused by the fact that 
the interactions around the boundaries between the blocks
and the single sites such as
$H^{B_L-C_1}$ and $H^{C_2-B_R}$ are represented 
in the product space of the block and single site, and 
they are not renormalized to reproduce
the ground-state wavefunction within a small number of basis states.  
When the interactions across the boundaries are weak,
the accuracy of the DMRG calculation will be improved.

\section{Finite temperature DMRG}

The zero-temperature DMRG method described in the above section
enables us to calculate 
the ground state wavefunction of large-size of systems 
with controlled accuracy. 
However, in order to study the properties of the bulk system, 
we need to extend the system in real space direction and 
analyze the asymptotic behaviors. 
In the finite-temperature DMRG method, 
the thermodynamic quantities of an infinitely long system are
directly obtained from the eigenvalues of QTM,
and temperature dependence is obtained in the process of
iterative extensions of the QTM in the $\beta$-direction.
In this section we describe how we calculate thermodynamic 
quantities using the DMRG method applied to the QTM.

\subsection{Transfer matrix method}
The thermodynamic quantities are calculated from the
partition function of the system.
In one-dimensional quantum systems the partition function is
obtained from the maximum eigenvalue of the
QTM\cite{Betsuyaku,Betsuyaku2}.
Here we first briefly summarize the way to calculate the QTM
and the partition function.

In the following we suppose that the Hamiltonian consists of only on-site
and nearest-neighbor interactions so that we can divide the Hamiltonian
into two parts
$H_{\mbox{\scriptsize odd}}=\sum_{n=0}^{L/2-1}h_{2n+1,2n+2}$ and 
$H_{\mbox{\scriptsize even}}=\sum_{n=0}^{L/2-1}h_{2n,2n+1}$.
Here $h_{i,i+1}$ contains 
on-site and nearest-neighbor interactions between the $i$th 
and $i$+1th sites and each $h_{i,i+1}$ commutes with all other 
$h_{i+2n,i+1+2n}$:
\begin{eqnarray}
[h_{2n,2n+1},h_{2n',2n'+1}] =
[h_{2n+1,2n+2},h_{2n'+1,2n'+2}]=0 .
\label{comute}
\end{eqnarray}
The partition function is then decomposed into a product of 
$e^{-\beta h_{i,i+1}/M}$ as 
\begin{eqnarray}
Z & = & {\mbox{Tr}}\ e^{-\beta H} \\ 
  & = & \lim_{M \rightarrow \infty} {\mbox{Tr}} 
           \left[ e^{-\beta H_{\mbox{\tiny odd}}/M}\ 
               e^{-\beta H_{\mbox{\tiny even}}/M}\right]^M \\
  & = & \lim_{M \rightarrow \infty} \mbox{Tr} 
         \left[\prod_{n=0}^{L/2-1} 
            e^{-\beta h_{2n+1,2n+2}/M}
         \prod_{n=0}^{L/2-1} 
            e^{-\beta h_{2n,2n+1}/M}\right]^M \label{Z}
\end{eqnarray}
where $M$ is the Trotter number\cite{Trotter,Suzuki1,Suzuki2}.

\begin{figure}
\epsfxsize=95mm \epsffile{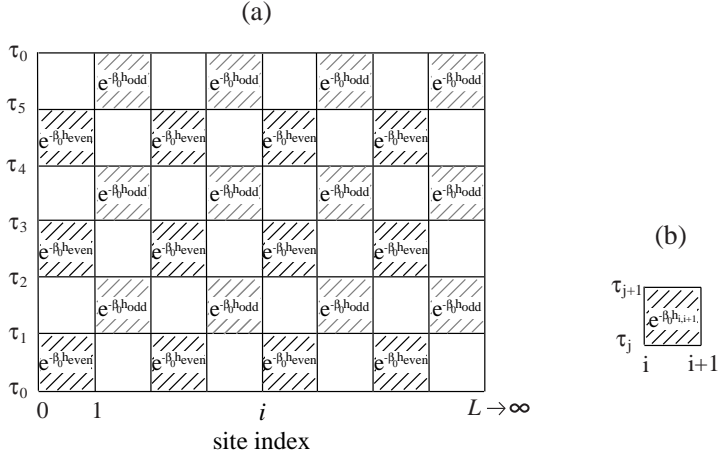}
\caption{\label{Fig_Z}
(a) Partition function $Z$ for $M=3$ represented by a tensor product of 
$e^{-\beta_0 h_{even}}$ and $e^{-\beta_0 h_{odd}}$,
where $h_{even}=h_{2n,2n+1}$ and $h_{odd}=h_{2n+1,2n+2}$.
(b) Graphical representation of 
$e^{-\beta_0 h_{i,i+1}}_{(\tau_{j+1},\tau_{j})}$ with $\beta_0=\beta/M$.
}
\end{figure}

Here we introduce imaginary time $\tau_j$, 
whose index $j$ takes only integers up to $2M$,
and we represent the partition function in a tensor product of 
$e^{-\beta h_{i,i+1}/M}$ under the complete basis set for $\tau_j$
as shown in figure \ref{Fig_Z}(a), where $e^{-\beta h_{i,i+1}/M}$ 
is represented by a hatched square shown in figure \ref{Fig_Z}(b).
Each $e^{-\beta h_{i,i+1}/M}$ is then represented by the basis states
specified by $i, i+1$ and $\tau_j, \tau_{j+1}$.
We explicitly write the imaginary time indices
$\tau_j$ and $\tau_{j+1}$ in $e^{-\beta h_{i,i+1}/M}$, and
define the tensor $e^{-\beta h_{i,i+1}/M}_{(\tau_{j+1},\tau_{j})}$ 
represented by
\begin{eqnarray}
e^{-\beta h_{i,i+1}/M}_{(\tau_{j+1},\tau_{j})} &\equiv& 
\sum_{\sigma_{i,\tau_{j+1}}}
\sum_{\sigma_{i+1,\tau_{j+1}}} 
\sum_{\sigma_{i,\tau_{j}}'}
\sum_{\sigma_{i+1,\tau_{j}}'} \nonumber \\
&&|\sigma_{i,\tau_{j+1}} \sigma_{i+1,\tau_{j+1}}\rangle 
\langle \sigma_{i} \sigma_{i+1}| e^{-\beta h_{i,i+1}/M} 
| \sigma_{i}' \sigma_{i+1}' \rangle 
\langle \sigma_{i,\tau_{j}}' 
\sigma_{i+1,\tau_{j}}'| 
\end{eqnarray}
where $\sigma_{i,\tau_{j}}$ is the local variable which specifies the state
at the $i$th site and at imaginary time $\tau_{j}$.

\begin{figure}
\epsfxsize=75mm \epsffile{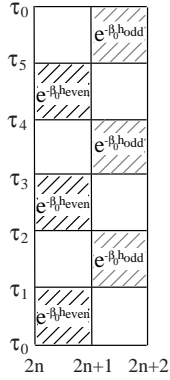}
\caption{\label{Fig_TM}
Quantum transfer matrix $\cal T$ represented by a tensor product of 
$e^{-\beta h_{i,i+1}/M}$ with $M=3$.
}
\end{figure}

The QTM ${\cal T}_n$ is written by
\begin{eqnarray}
{\cal T}_n & = & 
\lim_{M \rightarrow \infty}  
        \prod_{j=0}^{M-1}(
            e^{-\beta h_{2n+1,2n+2}/M}_{(\tau_{2j+2},\tau_{2j+1})} \ 
            e^{-\beta h_{2n,2n+1}/M}_{(\tau_{2j+1},\tau_{2j})}) \label{QTM}
\end{eqnarray}
as a tensor product of $e^{-\beta h_{i,i+1}/M}_{(\tau_{j+1},\tau_{j})}$
with the periodic boundary conditions in the $\beta$-direction,
$\tau_{2M}=\tau_{0}$.
This matrix ${\cal T}_n$ is graphically represented in figure \ref{Fig_TM}. 
When the system is translationally symmetric, we can omit the site 
index $n$ in ${\cal T}_n$, and the partition function is given by
\begin{eqnarray}
Z  & = & \mbox{Tr}\ {\cal T}^{L/2} .
\end{eqnarray}
Using the eigenvectors of the transfer matrix as basis states, 
we can represent the partition function using
their eigenvalues $\lambda_i$ as
\begin{eqnarray}
Z  & = & \sum_i \lambda_i^{L/2}.
\end{eqnarray}
In the thermodynamic limit $L\rightarrow \infty$, $Z$ is 
determined only by the maximum eigenvalue $\lambda_{max}$:
\begin{eqnarray}
Z=\lambda^{L/2}_{max}.
\end{eqnarray}
The free energy par site is then calculated from the  
maximum eigenvalue $\lambda_{max}$ as
\begin{eqnarray}
F & = & -\ln Z /(L \beta) \\
  & = & -\ln \lambda_{max}/(2\beta).
\end{eqnarray}
The thermodynamic quantities such as
the entropy, specific heat and the magnetic susceptibility 
are given by taking the derivative of $F$ with respect to
the temperature and magnetic field.

\subsection{Application of the DMRG to the quantum transfer matrix}
In order to calculate low temperature properties, we need to 
increase the Trotter number $M$ keeping $\beta/M=\Delta\tau$ 
sufficiently small.
However, the dimension of the QTM exponentially increases
with increasing the Trotter number.
In this section we describe how we apply the DMRG method to
QTM and obtain the maximum eigenvalue for large $M$.

In the zero-temperature DMRG method, we extend the Hamiltonian in 
the real space direction by restricting the number of basis states
by using the eigenvectors of the density matrix calculated from 
the ground-state wavefunction.
In the finite-temperature DMRG method, we extend 
the transfer matrix in the $\beta$-direction by restricting the 
basis states using the density matrix calculated from the 
eigenvector of the maximum eigenvalue $\lambda_{max}$. 

\begin{figure}
\epsfxsize=145mm \epsffile{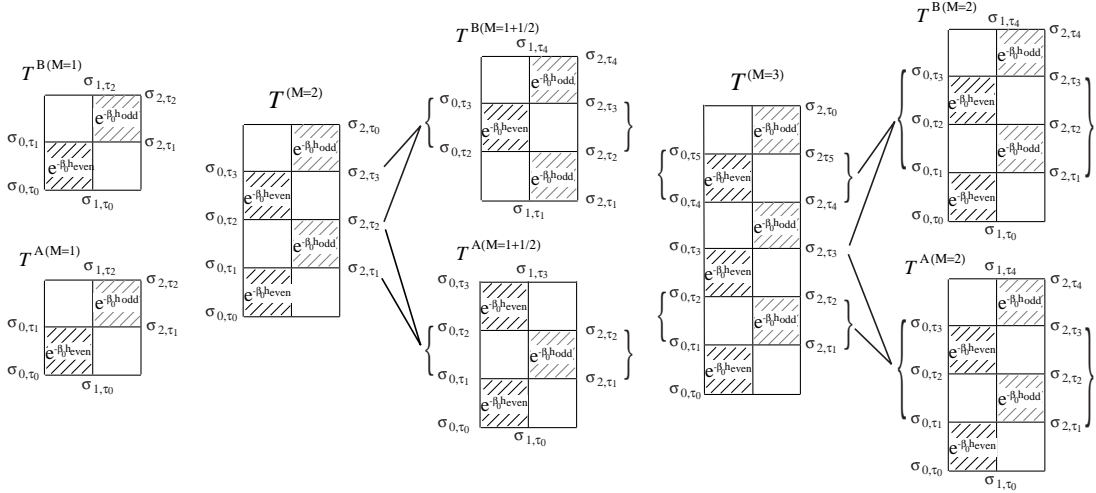}
\caption{\label{Fig_FTDMex}
Schematic diagram for the expansion of the QTM.
}
\end{figure}

We start from a small QTM with $M=2$ and divide it into two blocks, 
${\cal T}^{A(M=1)}$ and ${\cal T}^{B(M=1)}$, as
\begin{eqnarray}
\label{TMB1}
{\cal T}^{A(M=1)}_{(\tau_{2},\tau_{1},\tau_{0})}&=&
{\cal T}^{B(M=1)}_{(\tau_{2},\tau_{1},\tau_{0})} \nonumber \\
&=&
e^{-\beta_0 h_{\mbox{\tiny odd}}}_{(\tau_{2},\tau_{1})} \ 
e^{-\beta_0 h_{\mbox{\tiny even}}}_{(\tau_{1},\tau_{0})}
\end{eqnarray}
where $h_{\mbox{\scriptsize odd}}=h_{2n+1,2n+2}$ and
$h_{\mbox{\scriptsize even}}=h_{2n,2n+1} $.
The transfer matrix ${\cal T}$ is defined by
\begin{eqnarray}
\label{TM2}
{\cal T}^{(2M)}
&=& {\cal T}^{B(M)}_{(\tau_0,\tau_{(4M-1,\cdots,2M+2,2M+1)},\tau_{2M})}\ 
{\cal T}^{A(M)}_{(\tau_{2M},\tau_{(2M-1,\cdots,2,1)},\tau_0)}\\
{\cal T}^{(2M+1)}
&=& {\cal T}^{B(M+1/2)}_{(\tau_0,\tau_{(4M+1,\cdots,2M+3,2M+2)},\tau_{2M+1})}\ 
{\cal T}^{A(M+1/2)}_{(\tau_{2M+1},\tau_{(2M,\cdots,2,1)},\tau_0)}
\end{eqnarray}
depending on the parity of total Trotter number, and we have
imposed the periodic boundary conditions,
$\sigma_{i,\tau_{4M}}=\sigma_{i,\tau_0}$, in the $\beta$-direction.
We extend ${\cal T}^{A}$ and ${\cal T}^{B}$ as
\begin{eqnarray}
e^{-\beta_0 h_{\mbox{\tiny even}}}_{(\tau_{2M+1},\tau_{2M})} 
{\cal T}^{A(M)}_{(\tau_{2M},\tau_{(2M-1,\cdots,2,1)},\tau_{0})} \ 
& \rightarrow & 
{\cal T}^{A(M+1/2)}_{(\tau_{2M+1},\tau_{(2M,\cdots,2,1)},\tau_{0})} \\
{\cal T}^{B(M)}_{(\tau_{2M+2},\tau_{(2M+1,\cdots,4,3)},\tau_2)}\ 
e^{-\beta_{0} h_{\mbox{\tiny odd}}}_{(\tau_2,\tau_1)}  
& \rightarrow & 
{\cal T}^{B(M+1/2)}_{(\tau_{2M+2},\tau_{(2M+1,\cdots,3,2)},\tau_{1})} \\
e^{-\beta_0 h_{\mbox{\tiny odd}}}_{(\tau_{2M+2},\tau_{2M+1})} \
{\cal T}^{A(M+1/2)}_{(\tau_{2M+1},\tau_{(2M,\cdots,2,1)},\tau_{0})} 
& \rightarrow & 
{\cal T}^{A(M+1)}_{(\tau_{2M+2},\tau_{(2M+1,\cdots,2,1)},\tau_{0})} \\ 
{\cal T}^{B(M+1/2)}_{(\tau_{2M+2},\tau_{(2M+2,\cdots,3,2)},\tau_{1})} 
e^{-\beta_0 h_{\mbox{\tiny even}}}_{(\tau_1,\tau_0)} \ 
& \rightarrow & 
{\cal T}^{B(M+1)}_{(\tau_{2M+2},\tau_{(2M+1,\cdots,2,1)},\tau_{0})} 
\label{TMB_2}
\end{eqnarray} 
as shown in figure \ref{Fig_FTDMex}.

At each expansion of the QTM in the $\beta$-direction, 
we restrict the basis states of ${\cal T}^{A}$ and ${\cal T}^{B}$ 
while keeping the accuracy of $\lambda_{max}$ and corresponding
eigenvectors, which satisfy the following equations
\begin{eqnarray}
\label{TMEVeq1}
\sum_{j}\ V^L_{j} \ {\cal T}_{j j'} &=& \lambda_{max} V^L_{j'} \\
\label{TMEVeq2}
\sum_{j'} \ {\cal T}_{j j'} V^R_{j'} &=& \lambda_{max} V^R_{j}
\end{eqnarray}
where ${\cal T}_{j j'}$ represents the matrix element of ${\cal T}$
between the basis states $j$ and $j'$, 
and $V^L$ and $V^R$ are the left and the right eigenvectors
of the maximum eigenvalue $\lambda_{max}$, normalized by
$\langle V^L  | V^R \rangle \equiv 1$.
$V^L$ and $V^R$ are usually different because ${\cal T}$ is non-Hermitian.

Since the maximum eigenvalue $\lambda_{max}$ 
of ${\cal T}$ is equivalent to $\langle V^L | {\cal T} | V^R \rangle$,
we need to keep the norm $\langle V^L  | V^R \rangle$ 
as much as possible when we restrict the basis states.
To find the optimal basis states of ${\cal T}^{A}$,
we calculate the following density matrix defined by
\begin{eqnarray}
\rho_{k k'}^A = \sum_{l} V^L_{k l} V^R_{{k'} l}\ 
\end{eqnarray}
where index $k$ represents the basis states 
correspond to the imaginary time between $\tau_1$ and $\tau_{2M}$
(between $\tau_1$ and $\tau_{2M-1}$) for 
${\cal T}^{A(M+1/2)}$ (${\cal T}^{A(M)}$), 
which is transformed to new basis states,  
and $l$ represents the basis states for the remaining imaginary time: 
$\{j\}=\{k,l\}$.
Since $\langle V^L  | V^R \rangle$
is given by the sum of the eigenvalues $w^\alpha$ of the density matrix, 
$\langle V^L  | V^R \rangle =
 \mbox{Tr}\ \rho^A = \sum_k \rho_{k k}^A = \sum_\alpha w_\alpha$,
%important basis states 
%are obtained from the eigenvectors of large eigenvalues.
the optimal basis set within a fixed number $m$ of
basis states are obtained from the eigenvectors of the $m$ largest 
eigenvalues.
The new basis states are thus defined by
the eigenvectors $\vec{v}^{\alpha (L)}$ and $\vec{v}^{\alpha (R)}$ of 
the density matrix, which satisfy the eigenvalue equations
\begin{eqnarray}
\label{DMEV1}
\sum_{k}\ v^{\alpha (L)}_{k} \rho^A_{k k'} 
&=& w^\alpha v^{\alpha (L)}_{k'} \\
\label{DMEV2}
\sum_{k'}\ \rho^A_{k k'} \ v^{\alpha (R)}_{k'}   
&=& w^\alpha v^{\alpha (R)}_{k}.
\end{eqnarray}

By using $\vec{v}^{\alpha (L)}$ and $\vec{v}^{\alpha (R)}$
we transform ${\cal T}^A$ as
\begin{equation}
{\cal T}^{A}_{\alpha \alpha'}
 = \sum_{k k'} v^{\alpha (R)}_k {\cal T}^{A}_{k k'}
v^{\alpha' (L)}_{k'}.
\end{equation}
Here, we have omitted the indices of the basis states 
corresponding to the part of the Hilbert space with 
imaginary time $\tau_0$ and $\tau_{2M+1}$ ($\tau_0$ and $\tau_{2M}$)
for ${\cal T}^{A(M+1/2)}$ (${\cal T}^{A(M)}$),
which are included in ${\cal T}^{A}$ but 
independent of the transformation.
As a result of the non-Hermitian density matrix, the left and the right 
eigenvectors, $\vec{v}^{\alpha (L)}$ and $\vec{v}^{\alpha (R)}$, are different,
and this difference makes a different transformation between the left and 
right basis states of the transfer matrix.
Since the left and right eigenvectors 
$\vec{v}^{\alpha (L)}$ and $\vec{v}^{\alpha (R)}$ are dual orthogonal
\begin{equation}
 \sum_k v^{\alpha (L)}_k  v^{\alpha' (R)}_k = \delta_{\alpha \alpha'} 
\end{equation}
the orthogonality of the left and right basis states
is retained.

\begin{figure}
\epsfxsize=75mm \epsffile{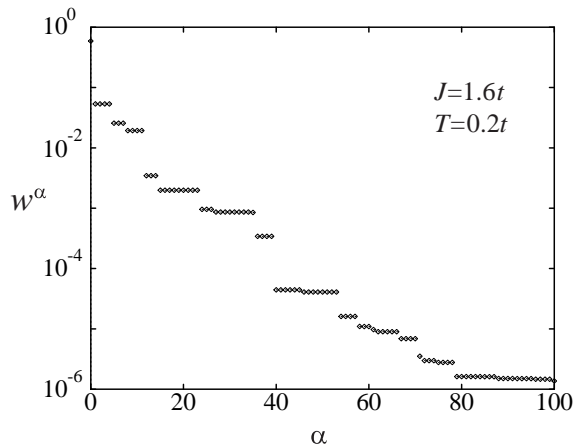}
\caption{\label{Fig_FTDM}
Eigenvalues of the density matrix $\rho$ calculated 
from the eigenvectors for $\lambda_{max}$ of the QTM
in one-dimensional Kondo lattice model at half-filling.}
\end{figure}

In this way, we restrict
the number of basis states while keeping the accuracy of the 
maximum eigenvalue and its eigenvectors.
The truncation error in the calculation is estimated from the
sum of the eigenvalues of the density matrix which are truncated off, 
and it is given by $1-\sum_{k=1}^m w^{k}$, where
$w^{k}$ is the $k$th largest eigenvalue.
In figure \ref{Fig_FTDM}
we show a typical example of the eigenvalues of the density matrix.
We can see only a small number of eigenstates have
large eigenvalues and the truncation error is about $10^{-4}$ for $m=50$.
In such a case, we can repeat the expansion of the QTM
and obtain its maximum eigenvalue with controlled accuracy.

\subsection{Quantum transfer matrix}
Here we explicitly write the QTM.
The QTM is obtained by extracting the matrix related to
the site indices $2n,2n+1$, and $2n+2$ 
from the partition function
decomposed into a tensor product of 
$e^{-\beta_0 h_{2n+1,2n+2}}_{(\tau_{2m+2},\tau_{2m+1})}$
and
$e^{-\beta_0 h_{2n,2n+1}}_{(\tau_{2m+1},\tau_{2m})}$.
We describe these local tensors using the local 
variables $\sigma_{i,\tau_j}$
at site $i$ and imaginary time $\tau_j$.
$\sigma$ represents up-spin state or down-spin state
for $S=1/2$ Heisenberg model. 
The transfer matrix ${\cal T}^{(M)}$ is then written by
\begin{eqnarray}
{\cal T}^{(M)} 
=\sum_{\sigma_{1,\tau_0}\cdots\sigma_{1,\tau_{2M-1}}}
\Theta (\sigma_{1,\tau_{0}}\sigma_{2,\tau_{0}}|
\sigma_{1,\tau_{2M-1}}\sigma_{2,\tau_{2M-1}}) \cdots \nonumber \\
%\Theta (\sigma_{1,\tau_{2}}\sigma_{2,\tau_{2}}|
%\sigma_{1,\tau_{1}}\sigma_{2,\tau_{1}})
\Theta (\sigma_{0,\tau_{3}}\sigma_{1,\tau_{3}}|
\sigma_{0,\tau_{2}}\sigma_{1,\tau_{2}}) 
\Theta (\sigma_{1,\tau_{2}}\sigma_{2,\tau_{2}}|
\sigma_{1,\tau_{1}}\sigma_{2,\tau_{1}})
\Theta (\sigma_{0,\tau_{1}}\sigma_{1,\tau_{1}}|
\sigma_{0,\tau_{0}}\sigma_{1,\tau_0})
\end{eqnarray}
where
\begin{equation}
\Theta (\sigma_{i,\tau_{j+1}}\sigma_{i+1,\tau_{j+1}}|
\sigma_{i,\tau_{j}}\sigma_{i+1,\tau_j})=
\langle\sigma_{i,\tau_{j+1}},\sigma_{i+1,\tau_{j+1}}|
e^{-\beta_0 h_{\mbox{\tiny i,i+1}}}_{(\tau_{j+1},\tau_{j})}|
\sigma_{i,\tau_{j}},\sigma_{i+1,\tau_{j}}\rangle
\end{equation}
and 
$\beta_0=\beta/M=2(\tau_{j+1}-\tau_{j})=\Delta\tau$.

The matrix elements 
$\Theta (\sigma_{i,\tau_{j+1}}\sigma_{i+1,\tau_{j+1}}|
\sigma_{i,\tau_{j}}\sigma_{i+1,\tau_j})$
are determined by using eigenstates of the local Hamiltonian $h_{i,i+1}$.
In the case of $S$=1/2 Heisenberg model,
$h_{i,i+1}$ is $J {\bf S}_i \cdot {\bf S}_{i+1}$.
The eigenvalues $E_m$ are $(-3J/4,J/4,J/4,J/4)$ and the corresponding
eigenvectors are
\begin{eqnarray}
|V_1\rangle&=&\frac{1}{\sqrt{2}}(|\uparrow,\downarrow\rangle - 
   |\downarrow,\uparrow\rangle) \\
|V_2\rangle&=&\frac{1}{\sqrt{2}}(|\uparrow,\downarrow\rangle + 
   |\downarrow,\uparrow\rangle) \\
|V_3\rangle&=&|\uparrow,\uparrow\rangle \\
|V_4\rangle&=&|\downarrow,\downarrow\rangle.
\end{eqnarray}
The matrix element 
$\langle \uparrow_{i},\downarrow_{i+1} |
e^{-\beta_0 h_{\mbox{\tiny i,i+1}}}| 
\uparrow_{i},\downarrow_{i+1} \rangle $
is then given by
\begin{eqnarray}
& \langle \uparrow_{i},\downarrow_{i+1} |
e^{-\beta_0 h_{\mbox{\tiny i,i+1}}}| 
 \uparrow_{i},\downarrow_{i+1} \rangle \nonumber \\
& =\sum_m \langle \uparrow_{i},\downarrow_{i+1}|V_m\rangle
\langle V_m |
e^{-\beta_0 E_m}| V_m\rangle \langle V_m |
 \uparrow_{i},\downarrow_{i+1} \rangle \nonumber\\
& = \frac{1}{2}(e^{\beta_0 3J/4}+e^{-\beta_0 J/4}) .
\end{eqnarray}
Similarly, off-diagonal elements
and other diagonal elements are obtained as follows:
\begin{eqnarray}
 \langle \uparrow_{i},\downarrow_{i+1} |
e^{-\beta_0 h_{\mbox{\tiny i,i+1}}}| 
 \downarrow_{i},\uparrow_{i+1} \rangle 
& = &
 \langle \downarrow_{i},\uparrow_{i+1} |
e^{-\beta_0 h_{\mbox{\tiny i,i+1}}}| 
 \uparrow_{i},\downarrow_{i+1} \rangle \nonumber\\
&=& -\frac{1}{2}(e^{\beta_0 3J/4}-e^{-\beta_0 J/4}) \\
 \langle \uparrow_{i},\uparrow_{i+1} |
e^{-\beta_0 h_{\mbox{\tiny i,i+1}}}| 
 \uparrow_{i},\uparrow_{i+1} \rangle 
& = & 
 \langle \downarrow_{i},\downarrow_{i+1} |
e^{-\beta_0 h_{\mbox{\tiny i,i+1}}}| 
 \downarrow_{i},\downarrow_{i+1} \rangle \nonumber\\
&=& e^{-\beta_0 J/4} .
\end{eqnarray}
Thus, the matrix 
$\Theta (\sigma_{i,\tau_{j+1}}\sigma_{i+1,\tau_{j+1}}|
\sigma_{i,\tau_{j}}\sigma_{i+1,\tau_j})$
is represented by
\begin{equation} 
\label{TMel}
e^{\beta_0 J/4}\left(
\begin{array}{cccc}
e^{-\beta_0 J/2}&0&0&0\\
0&\cosh(-\beta_0 J/2)&
\sinh(-\beta_0 J/2)&0\\
0&\sinh(-\beta_0 J/2)&
\cosh(-\beta_0 J/2)&0\\
0&0&0&e^{-\beta_0 J/2}
\end{array}
\right)
\end{equation}
with the basis states of the matrix
\[
(\,|\!\uparrow_{i,\tau_{j}},\uparrow_{i+1,\tau_{j}} \rangle , 
\,|\!\uparrow_{i,\tau_{j}},\downarrow_{i+1,\tau_{j}} \rangle,
\,|\!\downarrow_{i,\tau_{j}},\uparrow_{i+1,\tau_{j}} \rangle,
\,|\!\downarrow_{i,\tau_{j}},\downarrow_{i+1,\tau_{j}} \rangle\,).
\]

In order to explicitly represent the QTM,
we need to use the basis states 
\[
(\,|\!\uparrow_{i,\tau_{j}},\uparrow_{i,\tau_{j+1}} \rangle ,
\,|\!\uparrow_{i,\tau_{j}},\downarrow_{i,\tau_{j+1}} \rangle,
\,|\!\downarrow_{i,\tau_{j}},\uparrow_{i,\tau_{j+1}} \rangle,
\,|\!\downarrow_{i,\tau_{j}},\downarrow_{i,\tau_{j+1}} \rangle\,)
\]
defined along the $\beta$-axis at a given site $i$.
The matrix $\hat{\Theta}$ defined in these basis states is
related to the matrix $\Theta$ as
\begin{equation}
\hat{\Theta} (\sigma_{i,\tau_{j}}\sigma_{i,\tau_{j+1}}|
\sigma_{i+1,\tau_{j}}\sigma_{i+1,\tau_{j+1}} )=
\Theta (\sigma_{i,\tau_{j+1}}\sigma_{i+1,\tau_{j+1}}|
\sigma_{i,\tau_{j}}\sigma_{i+1,\tau_j}) .
\end{equation}
Thus, the matrix $\hat{\Theta}$ is represented by
\begin{equation}
\label{TMe}
e^{\beta_0 J/4}\left(
\begin{array}{cccc}
e^{-\beta_0 J/2}&0&0&\cosh(-\beta_0 J/2)\\
0&0&
\sinh(-\beta_0 J/2)&0\\
0&\sinh(-\beta_0 J/2)&
0&0\\
\cosh(-\beta_0 J/2)&0&0&e^{-\beta_0 J/2}
\end{array}
\right).
\end{equation}
This is the smallest unit of the QTM shown in figure \ref{Fig_Z}(b).

The initial blocks of the transfer matrix ${\cal T}^{A(M=1)}$
and ${\cal T}^{B(M=1)}$ in equation (\ref{TMB1}) are written as
\begin{eqnarray}
\label{TMAB}
{\cal T}^{A(M=1)}(\sigma_{0,\tau_0},\sigma_{0,\tau_{1}}|
\sigma_{1,\tau_0},\sigma_{1,\tau_{2}}|
\sigma_{2,\tau_1},\sigma_{2,\tau_{2}})  \nonumber \\
={\cal T}^{B(M=1)}(\sigma_{0,\tau_0},\sigma_{0,\tau_{1}}|
\sigma_{1,\tau_0},\sigma_{1,\tau_{2}}|
\sigma_{2,\tau_1},\sigma_{2,\tau_{2}})  \nonumber \\
=\sum_{\sigma_{1,\tau_1}}
\hat{\Theta} (\sigma_{0,\tau_{0}}\sigma_{0,\tau_1}|
\sigma_{1,\tau_{0}}\sigma_{1,\tau_{1}}) 
\hat{\Theta} (\sigma_{1,\tau_{1}}\sigma_{1,\tau_2}|
\sigma_{2,\tau_{1}}\sigma_{2,\tau_{2}}) .
\end{eqnarray}
The QTM ${\cal T}^{(M=1)}$
is obtained from ${\cal T}^{A(M=1)}$
with the periodic boundary conditions in the $\beta$-direction
$\sigma_{i,\tau_{2}}=\sigma_{i,\tau_{0}}$:
\begin{eqnarray}
\label{TM1}
{\cal T}^{(M=1)}(\sigma_{0,\tau_0},\sigma_{0,\tau_{1}}|
\sigma_{2,\tau_0},\sigma_{2,\tau_{1}})  \nonumber \\
=\sum_{\sigma_{1,\tau_0}}\sum_{\sigma_{1,\tau_2}}\sum_{\sigma_{2,\tau_2}}
{\cal T}^{A(M=1)}(\sigma_{0,\tau_0},\sigma_{0,\tau_{1}}|
\sigma_{1,\tau_0},\sigma_{1,\tau_{2}}|
\sigma_{2,\tau_1},\sigma_{2,\tau_{2}})  
\prod_{i=1}^2\delta_{\sigma_{i,\tau_{2}} \sigma_{i,\tau_{0}}}.
\end{eqnarray}

\subsection{Renormalization of the quantum transfer matrix}
The QTM at low temperature is obtained
by extending the QTM in the $\beta$-direction.
By using ${\cal T}^{A(M=1)}$ and ${\cal T}^{B(M=1)}$ defined
in equation (\ref{TMAB}), we first calculate the QTM of $M=2$.
This QTM corresponds to a high temperature of
$T=1/(M \beta_0) =1/(2 \beta_0)\gg 1$ since we need to
satisfy $\beta_0 =\Delta \tau \ll 1$. 
We therefore decrease the temperature by iteratively 
increasing Trotter number
$M$ as $T=1/(M \beta_0) \rightarrow T'=1/((M+1)\beta_0)$
keeping $\beta_0$ fixed.
In the following we describe detailed calculation for the 
extension of the QTM in the $\beta$-direction.

As is shown in figure \ref{Fig_FTDMex} 
the ${\cal T}^{(M=2)}$ is written 
by ${\cal T}^{A(M=1)}$ and ${\cal T}^{B(M=1)}$ as
\begin{eqnarray}
{\cal T}^{(M=2)}(\sigma_{0,\tau_0},\sigma_{0,\tau_{1}},
\sigma_{0,\tau_2},\sigma_{0,\tau_{3}}|
\sigma_{2,\tau_0},\sigma_{2,\tau_{1}},
\sigma_{2,\tau_2},\sigma_{2,\tau_{3}}) \nonumber \\
\mbox{\hspace{1cm}}=\sum_{\sigma_{1,\tau_0},\sigma_{1,\tau_2}}
{\cal T}^{B(M=1)}(\sigma_{0,\tau_2},\sigma_{0,\tau_{3}}|
\sigma_{1,\tau_2},\sigma_{1,\tau_{0}}|
\sigma_{2,\tau_3},\sigma_{2,\tau_{0}})  \nonumber \\
\mbox{\hspace{3cm}}\times{\cal T}^{A(M=1)}
(\sigma_{0,\tau_0},\sigma_{0,\tau_{1}}|
\sigma_{1,\tau_0},\sigma_{1,\tau_{2}}|
\sigma_{2,\tau_1},\sigma_{2,\tau_{2}}) . 
\end{eqnarray}
We first calculate the maximum eigenvalue $\lambda_{max}^{(M=2)}$
and the corresponding left and right eigenvectors, 
$V^{L(M=2)}(\sigma_{\tau_0},\sigma_{\tau_{1}},
\sigma_{\tau_2},\sigma_{\tau_{3}})$ and 
$V^{R(M=2)}(\sigma_{\tau_0},\sigma_{\tau_{1}},
\sigma_{\tau_2},\sigma_{\tau_{3}})$,
by solving eigenvalue equations 
(\ref{TMEVeq1}) and (\ref{TMEVeq2}).
We then calculate the density matrix $\rho$ for
the restriction of the basis states 
of ${\cal T}^{A(M=1+1/2)}$ and ${\cal T}^{B(M=1+1/2)}$:
\begin{eqnarray}
\rho(\sigma_{\tau_1},\sigma_{\tau_{2}}|
\sigma_{\tau_1}',\sigma_{\tau_{2}}') \nonumber \\
\mbox{\hspace{1cm}}=\sum_{\sigma_{\tau_0},\sigma_{\tau_3}}
V^{L(M=2)}(\sigma_{\tau_0},\sigma_{\tau_{1}},
\sigma_{\tau_2},\sigma_{\tau_{3}})
V^{R(M=2)}(\sigma_{\tau_0},\sigma_{\tau_{1}}',
\sigma_{\tau_2}',\sigma_{\tau_{3}}) \\
\rho(\sigma_{\tau_2},\sigma_{\tau_{3}}|
\sigma_{\tau_2}',\sigma_{\tau_{3}}') \nonumber \\
\mbox{\hspace{1cm}}=\sum_{\sigma_{\tau_0},\sigma_{\tau_1}}
V^{L(M=2)}(\sigma_{\tau_0},\sigma_{\tau_{1}},
\sigma_{\tau_2},\sigma_{\tau_{3}})
V^{R(M=2)}(\sigma_{\tau_0},\sigma_{\tau_{1}},
\sigma_{\tau_2}',\sigma_{\tau_{3}}').
\end{eqnarray}
The left and right eigenvectors 
 $v_\alpha^L(\sigma_{\tau_1},\sigma_{\tau_{2}})$ 
and $v_\alpha^R(\sigma_{\tau_1},\sigma_{\tau_{2}})$
of the density matrix 
$\rho(\sigma_{\tau_1},\sigma_{\tau_{2}}|
\sigma_{\tau_1}',\sigma_{\tau_{2}}') $
are determined by the eigenvalue equations (\ref{DMEV1}) and 
(\ref{DMEV2}). ${\cal T}^{A(M=1+1/2)}$ in the new 
basis states is obtained 
by using the eigenvectors of the $m$ largest eigenvalues of the
density matrix, $v_{\alpha_0}^R$ and $v_{\alpha_2}^L$ as
\begin{eqnarray}
{\cal T}^{A(M=1+1/2)}(\sigma_{0,\tau_0},\alpha_{0,\tau_{1,2}}
,\sigma_{0,\tau_{3}}|
\sigma_{1,\tau_0},\sigma_{1,\tau_{3}}|
\alpha_{2,\tau_{1,2}}) \nonumber \\
=\sum_{\sigma_{0,\tau_1},\sigma_{0,\tau_{2}}}
\sum_{\sigma_{2,\tau_1},\sigma_{2,\tau_{2}}}
\sum_{\sigma_{1,\tau_2}}
{\cal T}^{A(M=1)}(\sigma_{0,\tau_0},\sigma_{0,\tau_{1}}|
\sigma_{1,\tau_0},\sigma_{1,\tau_{2}}|
\sigma_{2,\tau_1},\sigma_{2,\tau_{2}})  \nonumber \\
\mbox{\hspace{1cm}}\times\hat{\Theta} (\sigma_{0,\tau_{2}}\sigma_{0,\tau_3}|
\sigma_{1,\tau_{2}}\sigma_{1,\tau_{3}}) 
v_{\alpha_0}^R(\sigma_{0,\tau_1},\sigma_{0,\tau_{2}})
v_{\alpha_2}^L(\sigma_{2,\tau_1},\sigma_{2,\tau_{2}}) 
\end{eqnarray}
where $\alpha_0$ and $\alpha_2$ corresponds
to the new basis states defined by the $m$ largest eigenvalues of the 
density matrix.
${\cal T}^{B(M=1+1/2)}$ is similarly given by
\begin{eqnarray}
{\cal T}^{B(M=1+1/2)}(\alpha_{0,\tau_{2,3}}|
\sigma_{1,\tau_1},\sigma_{1,\tau_{4}}|
\sigma_{2,\tau_1},\alpha_{2,\tau_{2,3}},\sigma_{2,\tau_{4}}) \nonumber \\
=\sum_{\sigma_{0,\tau_2},\sigma_{0,\tau_{3}}}
\sum_{\sigma_{2,\tau_2},\sigma_{2,\tau_{3}}}
\sum_{\sigma_{1,\tau_2}}
{\cal T}^{B(M=1)}(\sigma_{0,\tau_2},\sigma_{0,\tau_{3}}|
\sigma_{1,\tau_2},\sigma_{1,\tau_{4}}|
\sigma_{2,\tau_3},\sigma_{2,\tau_{4}})  \nonumber \\
\mbox{\hspace{1cm}}\times\hat{\Theta} (\sigma_{1,\tau_{1}}\sigma_{1,\tau_2}|
\sigma_{2,\tau_{1}}\sigma_{2,\tau_{2}}) 
v_{\alpha_0}^R(\sigma_{0,\tau_2},\sigma_{0,\tau_{3}})
v_{\alpha_2}^L(\sigma_{2,\tau_2},\sigma_{2,\tau_{3}}) .
\end{eqnarray}
The QTM of $M=3$ is obtained by using 
${\cal T}^{A(M=1+1/2)}$ and ${\cal T}^{B(M=1+1/2)}$ as
\begin{eqnarray}
{\cal T}^{(M=3)}(\sigma_{0,\tau_0},\alpha_{0,\tau_{1,2}},
\sigma_{0,\tau_3},\alpha_{0,\tau_{4,5}}|
\sigma_{2,\tau_0},\alpha_{2,\tau_{1,2}},
\sigma_{2,\tau_3},\alpha_{2,\tau_{4,5}}) \nonumber \\
=\sum_{\sigma_{1,\tau_0},\sigma_{1,\tau_3}}
{\cal T}^{B(M=1+1/2)}(\alpha_{0,\tau_{4,5}}|
\sigma_{1,\tau_3},\sigma_{1,\tau_{0}}|
\sigma_{2,\tau_3},\alpha_{2,\tau_{4,5}},\sigma_{2,\tau_{0}}) \nonumber \\
\mbox{\hspace{1.5cm}}\times
{\cal T}^{A(M=1+1/2)}(\sigma_{0,\tau_0},\alpha_{0,\tau_{1,2}}
,\sigma_{0,\tau_{3}}|
\sigma_{1,\tau_0},\sigma_{1,\tau_{3}}|
\alpha_{2,\tau_{1,2}}) .
\end{eqnarray}
The density matrices needed for the calculation of
${\cal T}^{A(M=2)}$ and ${\cal T}^{B(M=2)}$ 
are calculated from the eigenvectors of the QTM of $M=3$:
\begin{eqnarray}
\rho(\alpha_{\tau_{1,2}},\sigma_{\tau_3}|
\alpha_{\tau_{1,2}}',\sigma_{\tau_3}') \nonumber \\
\mbox{\hspace{0cm}}=\sum_{\sigma_{\tau_0},\alpha_{\tau_{4,5}}}
V^{L(M=3)}(\sigma_{\tau_0},\alpha_{\tau_{1,2}},
\sigma_{\tau_3},\alpha_{\tau_{4,5}})
V^{R(M=3)}(\sigma_{\tau_0},\alpha_{\tau_{1,2}}',
\sigma_{\tau_3}',\alpha_{\tau_{4,5}}) \\
\rho(\sigma_{\tau_3},\alpha_{\tau_{4,5}}|
\sigma_{\tau_3}',\alpha_{\tau_{4,5}}') \nonumber \\
\mbox{\hspace{0cm}}=\sum_{\sigma_{\tau_0},\alpha_{\tau_{1,2}}}
V^{L(M=3)}(\sigma_{\tau_0},\alpha_{\tau_{1,2}},
\sigma_{\tau_3},\alpha_{\tau_{4,5}})
V^{R(M=3)}(\sigma_{\tau_0},\alpha_{\tau_{1,2}},
\sigma_{\tau_3}',\alpha_{\tau_{4,5}}').
\end{eqnarray}
${\cal T}^{A(M=2)}$ and ${\cal T}^{B(M=2)}$ are obtained 
by using the eigenvectors of the $m$ largest eigenvalues of the
density matrix:
\begin{eqnarray}
{\cal T}^{A(M=2)}(\sigma_{0,\tau_0},\alpha_{0,\tau_{1,2,3}}|
\sigma_{1,\tau_0},\sigma_{1,\tau_{4}}|
\alpha_{2,\tau_{1,2,3}},\sigma_{2,\tau_{4}}) \nonumber \\
=\sum_{\alpha_{0,\tau_{1,2}},\sigma_{0,\tau_{3}}}
\sum_{\alpha_{2,\tau_{1,2}},\sigma_{2,\tau_{3}}}
\sum_{\sigma_{1,\tau_3}}
{\cal T}^{A(M=1+1/2)}(\sigma_{0,\tau_0},\alpha_{0,\tau_{1,2}}
,\sigma_{0,\tau_{3}}|
\sigma_{1,\tau_0},\sigma_{1,\tau_{3}}|
\alpha_{2,\tau_{1,2}}) \nonumber \\
\mbox{\hspace{1cm}}\times\hat{\Theta} (\sigma_{1,\tau_{3}}\sigma_{1,\tau_4}|
\sigma_{2,\tau_{3}}\sigma_{2,\tau_{4}}) 
v_{\alpha_0}^R(\alpha_{0,\tau_{1,2}},\sigma_{0,\tau_{3}})
v_{\alpha_2}^L(\alpha_{2,\tau_{1,2}},\sigma_{2,\tau_{3}})  \\
\nonumber \\
{\cal T}^{B(M=2)}(\sigma_{0,\tau_2},\alpha_{0,\tau_{3,4,5}}|
\sigma_{1,\tau_2},\sigma_{1,\tau_{6}}|
\alpha_{2,\tau_{3,4,5}},\sigma_{2,\tau_{6}}) \nonumber \\
=\sum_{\sigma_{0,\tau_{3}},\alpha_{0,\tau_{4,5}}}
\sum_{\sigma_{2,\tau_{3}},\alpha_{2,\tau_{4,5}}}
\sum_{\sigma_{1,\tau_3}}
{\cal T}^{B(M=1+1/2)}(\alpha_{0,\tau_{4,5}}|
\sigma_{1,\tau_3},\sigma_{1,\tau_{6}}|
\sigma_{2,\tau_3},\alpha_{2,\tau_{4,5}},\sigma_{2,\tau_{6}}) \nonumber \\
\mbox{\hspace{1cm}}\times\hat{\Theta} (\sigma_{0,\tau_{2}}\sigma_{0,\tau_3}|
\sigma_{1,\tau_{2}}\sigma_{1,\tau_{3}}) 
v_{\alpha_0}^R(\sigma_{0,\tau_{3}},\alpha_{0,\tau_{4,5}})
v_{\alpha_2}^L(\sigma_{2,\tau_{3}},\alpha_{2,\tau_{4,5}}) . \label{TMBexp_2}
\end{eqnarray}
${\cal T}^{B(M=2)}$ shown in figure~\ref{Fig_FTDMex} and 
equation (\ref{TMB_2}) is obtained
by shifting the imaginary time indices as $\tau_i \rightarrow \tau_{i-2}$.
Repeating the above procedures, we extend the QTM in the $\beta$-direction,
and the QTM of large $M$ is obtained within a restricted number 
of basis states.

\subsection{Thermodynamic quantities}
The iterative extension of the QTM in the $\beta$-direction
corresponds to a successive decrease in 
temperature, which is related to the Trotter number as 
%\begin{equation}
$T=1/\beta=1/(\beta_0 M)$.
%\end{equation}
At each temperature, the free energy of the system par site 
is obtained from the maximum eigenvalue $\lambda_{max}$
\begin{equation}
F(T)= -\ln \lambda_{max}/(2\beta).
\end{equation}
The entropy $S$ is obtained by taking derivative of the
free energy with respect to temperature
\begin{equation}
S(T)= -\frac{\partial}{\partial T} F .
\end{equation}
The specific heat $C$ is similarly obtained
by taking second derivative of the free energy
\begin{equation}
C(T)= -T\frac{\partial^2}{\partial T^2} F .
\end{equation}
The magnetic susceptibility $\chi_s$ and the charge susceptibility $\chi_c$ 
are also calculated from the free energy at slightly
different magnetic fields $h^z$ and chemical potentials $\mu$
\begin{eqnarray}
\chi_s&=&-\frac{\partial^2}{{\partial h^z}^2} F  \\ 
\chi_c&=&-\frac{\partial^2}{{\partial \mu}^2} F .
\end{eqnarray}

\begin{figure}
\epsfxsize=75mm \epsffile{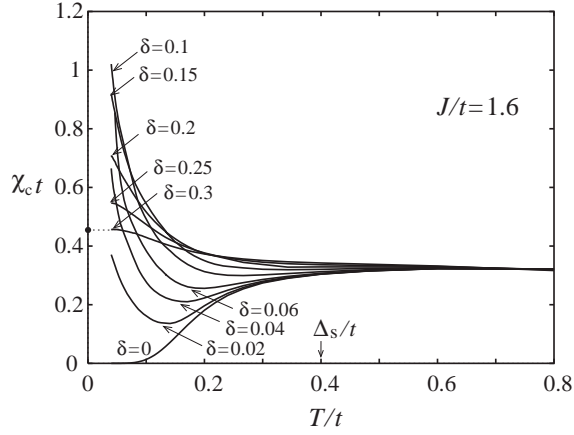}
\caption{\label{FigKLM_cc}
Charge susceptibility of the one-dimensional Kondo lattice model 
with hole doping $\delta$ 
in the insulating state at half-filling\cite{FT_KLM_d}.
The solid circle on the vertical axis represent the results
obtained by the original zero-temperature DMRG method.
$\Delta_s$ is the spin gap at $\delta=0$.
$J$ is the antiferromagnetic coupling between the conduction 
electrons and localized spins. $t$ is the hopping integral.
}
\end{figure}

The local quantities such as the magnetization and the 
density of electrons are obtained by using the 
eigenvectors $V^R$ and $V^L$ of $\lambda_{max}$.
The magnetization 
$\langle S^z_0 \rangle$ and the density
of electrons $\langle n_0 \rangle$ are given by
\begin{eqnarray}
\langle S^z_0 \rangle
&\equiv& \mbox{Tr}\{\mbox{e}^{-\beta H} S^z_0 \}/Z 
= \langle V^L | S^z_{0, \tau_0} | V^R \rangle  \\
\langle n_0 \rangle &\equiv& \mbox{Tr}\{\mbox{e}^{-\beta H} n_0 \}/Z 
= \langle V^L | n_{0, \tau_0} | V^R \rangle 
\end{eqnarray}
where $S^z_{0, \tau_0}$ is the operator at site $i=0$ and 
imaginary time $\tau_0$.
The magnetic susceptibility $\chi_s$ is then obtained by 
applying a small magnetic field $h^z$
\begin{equation}
\chi_s=\lim_{h^z\rightarrow 0} \langle S^z_0 \rangle/h^z .
\end{equation}
Similarly, the charge susceptibility $\chi_c$ is obtained by
changing chemical potential $\mu$
\begin{equation}
\chi_c=\lim_{\Delta \mu \rightarrow 0} 
(\langle n_0 \rangle_{\mu+\Delta\mu} - \langle n_0 \rangle_{\mu})/\Delta \mu .
\end{equation} 
As a typical example, the charge susceptibility of the one-dimensional 
Kondo lattice model calculated 
by the present method is shown in figure \ref{FigKLM_cc}.

The nearest-neighbor and next-nearest neighbor correlations are 
also obtained by
using the eigenvectors $V^R$ and $V^L$ and the QTM ${\cal T}$
\begin{eqnarray}
\langle S^z_0 S^z_1 \rangle 
&\equiv& \mbox{Tr}\{\mbox{e}^{-\beta H}  S^z_0 S^z_1 \}/Z 
= \langle V^L |  S^z_{0, \tau_0} 
{\cal T} S^z_{1, \tau_0}| V^R \rangle / \lambda_{max}  \label{FT_stcor}\\
\langle S^z_0 S^z_2 \rangle 
&\equiv& \mbox{Tr}\{\mbox{e}^{-\beta H}  S^z_0 S^z_2 \}/Z 
= \langle V^L |  S^z_{0, \tau_0} 
{\cal T} S^z_{2, \tau_0}| V^R \rangle / \lambda_{max} .
\end{eqnarray}
In equation (\ref{FT_stcor}) we have used $S^z_i$ commute with
the local Hamiltonian $h_{i,i+1}$.
The internal energy of the system is calculated from the 
above expectation values and we can calculate
the specific heat $C$ by taking 
derivative of the internal energy $E$ with respect to temperature
\begin{equation}
C(T)= \frac{\partial}{\partial T} E.
\end{equation}

In order to confirm the accuracy of the results, 
we need to increase 
the number of basis states $m$ kept in each block and
the Trotter number $M$ of the QTM. 
The error caused by finite $M$ is usually 
scaled by $M^{-2}$, and the correct value in the limit 
of $M\rightarrow \infty$ is obtained by the linear 
extrapolation with respect to $M^{-2}$.
The error caused by finite $m$ does not
follow a simple scaling form in general. 
Nevertheless, the truncation error in the norm of the eigenvectors 
$\langle V^L | V^R \rangle$ is given by 
 $1-\sum_{k=1}^m w^{k}$, where $w^{k}$ is the $k$th largest 
eigenvalue of the density matrix. The 
correction for the thermodynamic quantities 
will be estimated from the truncation error.

\begin{figure}
\epsfxsize=75mm \epsffile{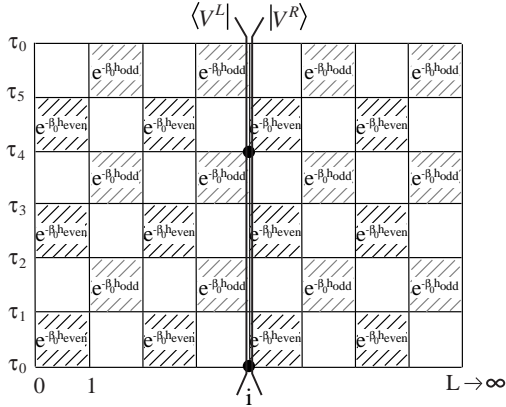}
\caption{\label{Fig_Dyn}
Schematic diagram of the calculation of imaginary-time correlation 
functions with the eigenvectors $\langle V^L |$ and $| V^R \rangle$.
}
\end{figure}

\subsection{Dynamic correlation function}
The Green's functions and dynamic correlation functions
are obtained from the imaginary time Green's functions 
and correlation functions through analytical continuation to
the real frequency axis\cite{Mutou}.
The imaginary time Green's function is calculated from
the eigenvectors of the QTM as
\begin{eqnarray}
G(\tau_{2j})
&\equiv& -\mbox{Tr}\{\mbox{e}^{-\beta H} c_{i \sigma}(\tau_{2j})\
c^\dagger_{i \sigma}(0)\}/Z \nonumber \\
&=& -\langle V^L| c_{i \sigma}(\tau_{2j})\ c^\dagger_{i \sigma}(0)
|V^R\rangle 
\label{G_tau}
\end{eqnarray}
where $\langle V^L | $ and $ | V^R \rangle$ are
eigenvectors of the largest eigenvalue $\lambda_{max}$,
and $\tau_{2j}=j(\beta/M)$.
The local dynamic correlation function $\chi_{AB}(\tau_{2j})$
on the $\beta$-axis is also obtained by
\begin{eqnarray}
\chi_{AB}(\tau_{2j})
%&=& \langle A^z(i\tau_{2j}) B^z(0)\rangle \nonumber \\
&\equiv& \mbox{Tr}\{\mbox{e}^{-\beta H} A_i(\tau_{2j})\ B_i(0)\}
/Z \nonumber \\
&=& \langle V^L|A_i(\tau_{2j})\ B_i(0)|V^R\rangle .
\end{eqnarray}
The frequency dependence of the imaginary time 
Green's function and correlation functions are obtained by
Fourier transformation along the imaginary axis
\begin{eqnarray}
G(i \omega_n) & = & \frac{\beta}{M}
\sum_{j=0}^M \mbox{e}^{i \omega_n \tau_{2j}} G(\tau_{2j}) \\
\chi_{AB}(i \omega_n) & = & \frac{\beta}{M}
\sum_{j=0}^M \mbox{e}^{i \omega_n \tau_{2j}} \chi_{AB}(\tau_{2j})
\end{eqnarray}
where $\omega_n$ is the Matsubara frequency that is
$\pi(2n+1)/\beta$ for fermionic operators and $2\pi n/\beta$
for bosonic operators.

The real frequency Green's functions and dynamic susceptibility
are obtained by the analytical continuation to the real frequency axis.
The Pad\'e approximations or the maximum entropy method\cite{MEM1,MEM2}
are used for this calculation. 
The Pad\'e approximation is a fitting of
$G(i \omega_n)$ or $\chi_{AB}(i \omega_n)$ by rational functions
of frequency $i \omega_n$, which are analytically continued to
the real axis by $i \omega_n \rightarrow \omega + i \delta$.

The maximum entropy method is based on the spectral representations
\begin{eqnarray}
G(\tau) &=& \int^{\infty}_{-\infty}
  \rho(\omega) \frac{\mbox{e}^{- \tau \omega}}{1+\mbox{e}^{- \beta \omega}
} d\omega  \\
\chi_{AB}(\tau) &=& \int^{\infty}_{-\infty}
\frac{1}{\pi}
\mbox{Im}\chi_{AB}(\omega)
\frac{\mbox{e}^{- \tau \omega}}{1-\mbox{e}^{- \beta \omega}} d\omega 
\end{eqnarray}
where $\rho(\omega)=-\frac{\mbox{Im}}{\pi}G(\omega+i\delta)$ is the 
density of state. 
This method finds the best $\rho(\omega)$ and $\chi_{AB}(\omega)$ 
that reproduce $G(\tau)$ and $\chi_{AB}(\tau)$.

The dynamic structure factor $S_{AB}(\omega)$ is obtained from
the imaginary part of $\chi_{AB}(\omega)$ through the fluctuation
dissipation theorem
\begin{equation}
\mbox{Im} \chi_{AB}(\omega) =
  \pi (1 - \mbox{e}^{-\beta \omega} ) S_{AB}(\omega).
\label{S_omega}
\end{equation}

\subsection{Finite-$T$ DMRG with fixed Trotter number}
\begin{figure}
\epsfxsize=75mm \epsffile{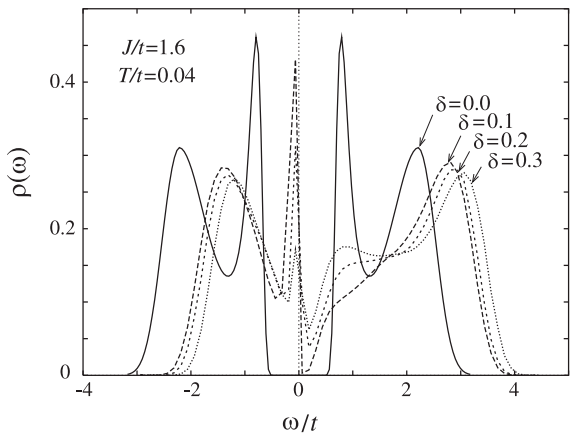}
\caption{\label{Fig_DynKLM}
Quasi-particle density of states calculated
in the one-dimensional Kondo lattice model with hole doping $\delta$ 
in the insulating state at half-filling\cite{Shibata_KLd}.
}
\end{figure}

Reliable analytical continuation is only possible when
we have accurate imaginary time correlation functions.
In order to obtain correct dynamic properties,
it is inevitable to use the finite system algorithm 
of the DMRG and to refine the eigenvectors $V^R$ and $V^L$
of the QTM.
In the finite system algorithm 
we reconstruct the blocks ${\cal T}^{A}$ 
and ${\cal T}^{B}$ by keeping the Trotter number fixed.
We first extend ${\cal T}^{A}$ until ${\cal T}^{B}$ is
reduced down to the initial block of ${\cal T}^{B(M=1)}$ , 
and then extend ${\cal T}^{B}$ to refine its basis states.
We repeat such sweeps until the eigenvectors $V^R$ and 
$V^L$ are converged.
Figure \ref{Fig_DynKLM} shows the quasi-particle density of states 
calculated from the eigenvectors obtained by the finite 
system algorithm of the DMRG.
In this figure we find the closing of the gap
at the Fermi level $\omega=0$ with the hole 
doping $\delta$ in the Kondo insulator\cite{Shibata_KLd}.

\subsection{Conserved quantities of the quantum transfer matrix}
The eigenvalues of the non-Hermitian matrix 
are not restricted to a real number. 
This fact sometimes causes uncontrolled instability
generated from small numerical errors.
In order to suppress such numerical instability and
to keep the accuracy of the calculation,
a precise diagonalization of the non-Hermitian matrix is needed.
For this purpose, it is efficient to reduce the Hilbert 
space by using the conserved quantities of the QTM,
which are related to the conservation law in the system.
For example, $S_i^z+S_{i+1}^z$ in the Heisenberg model is 
conserved at any imaginary time evolution generated by $h_{i,i+1}$.
This means the following equation is satisfied for each
imaginary time evolution from $\tau_{2k+i}$ to $\tau_{2k+i+1}$:
\begin{equation}
S_{i,\tau_{2k+i}}^z+S_{i+1,\tau_{2k+i}}^z = 
S_{i,\tau_{2k+i+1}}^z+S_{i+1,\tau_{2k+i+1}}^z .
\end{equation}
This equation is translated to
\begin{equation}
S_{i,\tau_{2k+i}}^z-S_{i,\tau_{2k+i+1}}^z = 
-S_{i+1,\tau_{2k+i}}^z+S_{i+1,\tau_{2k+i+1}}^z 
\end{equation}
and the following equation is obtained after the summation over $k$ 
\begin{equation}
\sum_{j=0}^{2M-1} (-1)^{j+i} S_{i,\tau_{j}}^z = \sum_{j=0}^{2M-1} (-1)^{j+i+1}
S_{i+1,\tau_{j}}^z.
\end{equation}
This result clearly shows that $Q_i$ defined by
\begin{equation}
Q_i=\sum_{j=0}^{2M-1} (-1)^{j+i} S_{i,\tau_{j}}^z 
\end{equation}
is a conserved quantity of the QTM\cite{Koma,Nomura}.

Similar calculations are possible for charge degrees of freedom.
For example,
\begin{equation}
P_i=\sum_{j=0}^{2M-1} (-1)^{j+i} n_{i,\tau_{j}} 
\end{equation}
is a conserved quantity of the QTM,
when $h_{i,i+1}$ preserves number of electrons $n_i+n_{i+1}$.

By using these conserved quantities, we can  
classify the basis states of the QTM.
Since there is no matrix element between different 
subspaces classified by $Q_i$ and $P_i$, the QTM becomes block diagonal.
The maximum eigenvalue $\lambda_{max}$ of the QTM 
is then determined by calculating the largest eigenvalue
in each subblock.
The  $\lambda_{max}$ of the QTM is obtained 
in the subblock of $Q_i=P_i=0$ in most cases.
The density matrix calculated from the eigenvectors 
of the QTM is also block diagonal, and
we need only diagonalize the density matrix within each subblock. 
The reduction of matrix dimension improves the accuracy
of the diagonalization and stabilizes the calculation.

\subsection{Complex eigenvalues of density matrix}
We finally comment on the problem of complex eigenvalues.
This is caused by the
non-Hermiticity of the density matrix, which
allows appearance of complex eigenvalues.
These are usually generated 
by the artificial truncation of basis states, and 
the imaginary parts of the eigenvalues are typically much 
smaller than their real parts.
The complex eigenvalues always appear with their complex conjugate, 
and their eigenvectors are mutually complex conjugate of the other one.
Thus we can define two real vectors, which span
the Hilbert space generated by the eigenvectors 
of the two complex eigenvalues,
from the real and the imaginary 
parts of the eigenvectors, 
provided that the imaginary parts of the eigenvalues are
sufficiently small compared with its real part.
After checking the dual orthogonality of the left and right
basis states $\langle i | j \rangle = \delta_{i j}$, 
we can continue the DMRG calculation 
using the basis states defined by the real vectors.

\section{Application to two-dimensional systems}
Electrons in two-dimensional systems 
exhibit various interesting phenomena. Although their properties 
have been extensively studied for a long time, many unresolved 
questions are still remaining. 
In order to study the ground-state and thermodynamic 
properties of the system, the quantum Monte Carlo method 
and the exact diagonalizations have been used.
However, when the electron density is away from particle hole 
symmetric point, the negative sign problem arises 
in quantum Monte Carlo simulations. 
Although the exact diagonalization method provides rigorous results, 
the number of electrons in the system is limited to small numbers. 
Thus, we need a new reliable method for two-dimensional electron 
systems away from half-filling, and 
various applications of the DMRG to two-dimensional systems have 
been considered.
Most of the applications use mappings on to  
effective one-dimensional systems.
However, the mapping is not uniquely determined and 
many long-range interactions appear, depending on the system.
We thus need to find appropriate mapping specific to each model.

\subsection{Mapping to the one-dimensional lattice model}
In this section we consider two-dimensional electrons in a magnetic field.
In two-dimensional systems under a perpendicular magnetic field, 
the kinetic energy of the electrons is completely 
quenched and the remaining 
macroscopic degeneracy is lifted by Coulomb interaction.
Such an interesting system is realized in quantum Hall systems,
where various electronic states have been observed.
Here we describe how the DMRG method is
applied to quantum Hall systems\cite{Shibata2,Shibata3}.

In order to represent the Hamiltonian in matrix form, we first 
define one-particle states  $\Psi_{N X}(x,y)$ 
of two-dimensional electrons.
We use the eigenstates of free electrons 
in perpendicular magnetic field and represent the wavefunction 
in the Landau gauge. 
The one-particle states are represented by the following
wavefunctions and each state is uniquely specified by only 
two numbers $N$ and $X$:
\begin{equation}
\label{BWF}
\Psi_{N X} = C_{N} \exp{\left[i {k_y y} -\frac{(x-X)^2}
{2\ell^2}\right]} H_N\left[\frac{x-X}{\ell}\right]
\end{equation}
where $N$ is the Landau level index and $X$ is the $x$-component
of the guiding center coordinates of an electron in cyclotron motion. 
$H_N$ are Hermite
polynomials and $C_{N}$ is the normalization constant.
$\ell=(\hbar c /e H)^{1/2}$ is the magnetic length, which is
the length-scale of the system.
The guiding center $X$ is related to the momentum $k_y$ 
through $X=k_y\ell^2$. 
Since $k_y$ is discretized under the periodic 
boundary conditions, the guiding center $X$ 
takes only discrete values
\begin{equation}
X_n=2\pi\ell^2 n/ L_y
\end{equation}
where $L_y$ is the length of the unit cell in the $y$-direction.
After we have fixed the Landau level index, each one-particle 
state is uniquely specified by the guiding center $X_n$.
Thus the Hamiltonian is mapped on to an effective 
one-dimensional lattice model.

The ground state of two-dimensional electrons 
in a perpendicular magnetic field is determined only by 
the Coulomb interaction
\begin{equation}
V(r)= \frac{e^2}{\epsilon r}.
\end{equation}
The Coulomb interaction generates correlations between 
the electrons and stabilize various electronic states
depending on the filling $\nu$ of Landau levels. 
When the magnetic field is strong enough so that 
the Landau level splitting is sufficiently
large compared with the typical Coulomb interaction
$e^2/(\epsilon \ell)$,
the electrons in fully occupied Landau 
levels are inert and the ground state is
determined only by the electrons in the top most
partially filled Landau level.

The Hamiltonian is then written by
\begin{equation}
\label{2DH}
H= S \sum_n c_n^\dagger c_n + 
\frac{1}{2}\sum_{n_1} \sum_{n_2} \sum_{n_3} \sum_{n_4} 
A_{n_1 n_2 n_3 n_4} c_{n_1}^\dagger  c_{n_2}^\dagger c_{n_3} c_{n_4}
\end{equation}
where we have imposed periodic boundary conditions in both
$x$- and $y$-directions, and
$S$ is the classical Coulomb energy of Wigner crystal
with a rectangular unit cell of $L_x \times L_y$\cite{QHHS}.
$c_n^\dagger$ is the creation operator of the electron represented 
by the wavefunction defined in equation (\ref{BWF}) with $X=X_n$. 
$A_{n_1 n_2 n_3 n_4}$ are the matrix elements of the Coulomb 
interaction defined by
\begin{eqnarray}
A_{n_1 n_2 n_3 n_4}&=&\delta'_{n_1+n_2,n_3+n_4}\frac{1}{L_xL_y}
\sum_{\bf q} \delta'_{n_1-n_4,q_yL_y/2\pi}\frac{2\pi e^2}{\epsilon q}
 \nonumber\\
&&{\mbox{\hspace{1cm}}}\times\left[L_N(q^2\ell^2/2)\right]^2
\exp{ \left[-\frac{q^2 \ell^2}{2}-i(n_1-n_3)\frac{q_xL_x}{M} \right] } 
\end{eqnarray}
where $L_N(x)$ are Laguerre polynomials with $N$ being the Landau level
index\cite{Yoshioka}. 
$\delta_{n_1,n_2}' = 1$ when $n_1=n_2 (\mbox{mod}\ M)$ with
$M$ being the number of one-particle states in the unit cell, 
which is given by the area of the unit cell
$2\pi M \ell^2=L_xL_y$.

In order to obtain the ground-state wavefunction
we apply the DMRG method described in section 2.
We start from a small-size system consisting of
only four one-particle states whose indices $n$
are 1, 2, $M-1$, and $M$, and we
calculate the ground-state wavefunction. We then construct
the left block containing local states of $n=1$ and 2, and the 
right block containing  $n=M-1$ and $M$ from the eigenvectors
of the density matrices which are calculated from the 
ground-state wavefunction. 
We then add two local states $n=3$ and $M-2$ between the
two blocks and repeat the above procedure until
$M$ local states are included in the system.
We then apply the finite system algorithm of the DMRG
to refine the ground-state wavefunction.
After we obtain the convergence, 
we calculate correlation functions.

The ground-state pair correlation function $g({\bf r})$ in 
guiding center coordinates is defined by 
\begin{equation}
g({\bf r})=\frac{L_xL_y}{N_e(N_e-1)}
\langle \Psi | \sum_{i\ne j} \delta({\bf r-R}_i+{\bf R}_j) | \Psi \rangle 
\end{equation}
where ${\bf R}_i$ is the guiding center coordinate of the $i$th
electron, and it is calculated from the following equation
\begin{eqnarray}
g({\bf r})&=& 
\frac{1}{N_e(N_e-1)}\sum_{\bf q}\sum_{n_1,n_2,n_3,n_4} \exp
\left[ i{\bf q \cdot r}-\frac{q^2\ell^2}{2}-i(n_1-n_3)\frac{q_xL_x}{M}
\right] \nonumber\\
&& \mbox{\hspace{4cm}}\times \delta'_{n_1-n_4,q_yL_y/2\pi} \langle 
\Psi |  c_{n_1}^\dagger  c_{n_2}^\dagger c_{n_3} c_{n_4} | \Psi \rangle
\end{eqnarray}
where $\Psi$ is the ground state and $N_e$ is the total number of 
electrons.

\begin{figure}
\epsfxsize=75mm \epsffile{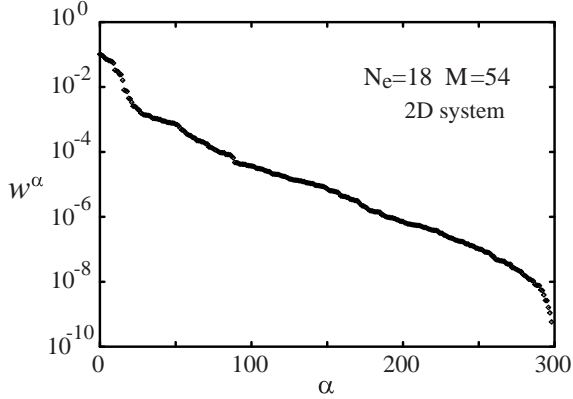}
\caption{\label{Fig_2DDM}
Eigenvalues of the density matrix for two-dimensional system with 18 electrons
in 54 local orbitals.}
\end{figure}

The accuracy of the results depends on the 
distribution of eigenvalues of the density matrix.
A typical example of the eigenvalues of the 
density matrix for system of $M=54$ with $18$ electrons
is shown in figure \ref{Fig_2DDM},
which shows an exponential decrease of eigenvalues $w^\alpha$.
In this case the accuracy of $10^{-4}$ is obtained
by keeping only 200 states in each block. 

\subsection{Conserved quantities in magnetic field} 
The Hamiltonian written in equation (\ref{2DH}) conserves
center of mass of electrons $G=\sum_i ({X_n})_i$,
whose origin is the conservation of total $y$-momentum
$\sum_i ({k_y})_i = \sum_i ({X_n})_i /\ell^2$.
All the eigenstates of the Hamiltonian and the 
basis states of the blocks are classified by
the quantum numbers, $N_e$ and $G$.
These conservation laws restrict 
the basis states of the ground state, 
and the complete basis states of the extended system
are not generally constructed 
from the eigenstates of the density matrix calculated
from the ground-state wavefunction which has fixed
numbers $N_e$ and $G$.
Although this unexpected truncation is corrected by 
reconstructing the blocks using the finite system algorithm 
of the DMRG, we need to include additional basis states 
in the extended blocks to accelerate convergence in the 
infinite system algorithm of the DMRG. 

\begin{figure}
\epsfxsize=75mm \epsffile{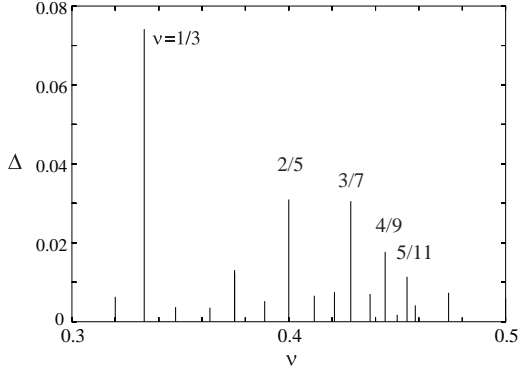}
\caption{\label{Fig_Gap}
The lowest excitation gap at various $\nu$ in 
the lowest Landau level. 
Relatively large excitation gap is obtained at
fractional fillings $\nu=n/(2n+1)$.
The excitation gap is in units of $e^2/(\epsilon \ell)$.
}
\end{figure}
\begin{figure}
\epsfxsize=75mm \epsffile{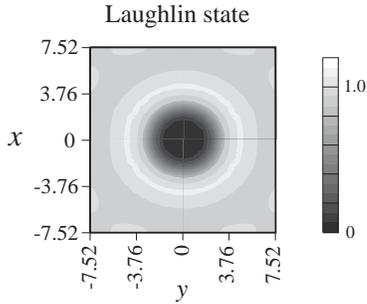}
\caption{\label{Fig_Lau}
Pair correlation function $g({\bf r})$ at $\nu=1/3$ in the lowest 
Landau level. The length is in units of $\ell$.
}
\end{figure}

\subsection{Diverse ground states in a magnetic field} 
Here we present diverse ground states obtained by the  
DMRG method applied to the quantum Hall systems.
In the limit of a strong magnetic field, the electrons occupy only
the lowest Landau level $N=0$. In this limit,
fractional quantum Hall effect (FQHE) has been 
observed at various fillings\cite{FQHEex}.
The FQHE state is characterized by incompressible liquid 
with a finite excitation gap\cite{Laugh}.
The presence of the  FQHE is confirmed by the DMRG calculations,
where a relatively large excitation gap is obtained at
various fillings between $\nu=1/2$ and 3/10\cite{Shibata3}
as shown in figure \ref{Fig_Gap}.
We clearly find large excitation gaps at fractional 
fillings $\nu=1/3,2/5,3/7,4/9$ and $5/11$,
which correspond to primary series of the FQHE at 
$\nu=n/(2n+1)$. 
The pair correlation function at $\nu=1/3$ is 
presented in figure \ref{Fig_Lau}, which shows a
circularly symmetric liquid state consistent with the 
Laughlin state\cite{Laugh}.

\begin{figure}
\epsfxsize=95mm \epsffile{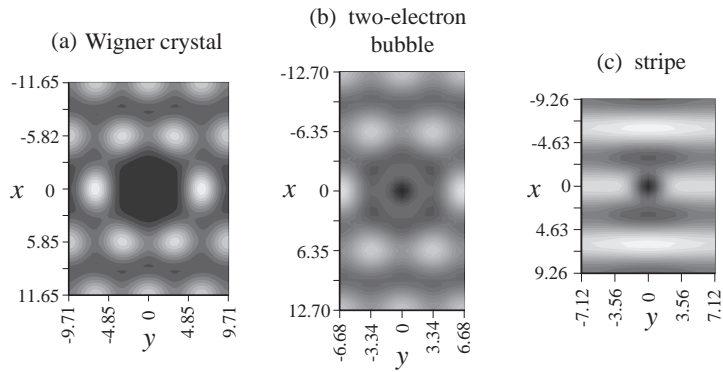}
\caption{\label{Fig_CDW}
Pair correlation functions $g({\bf r})$ in guiding center coordinates.
(a) The Wigner crystal realized in an excited state at $\nu=1/6$ in 
the lowest Landau level. The number of electrons in the unit cell $N_e$ is 12.
(b) Two-electron bubble state at $\nu=8/27$ in the third 
lowest Landau level. $N_e=16$.
(c) Stripe state at $\nu=3/7$ in the third lowest Landau level. $N_e=18$.
}
\end{figure}

In the limit of low filling $\nu\rightarrow 0$, mean 
separation between the electrons becomes much longer than the 
typical length-scale of the one-particle wavefunction.
In this limit the quantum fluctuations are not important 
and electrons behave as classical point particles,
whose ground state is the Wigner crystal. 
The formation of the Wigner crystal is also confirmed by 
the DMRG calculations at low fillings
as shown in figure \ref{Fig_CDW} (a).
The $\nu$ dependence of the low-energy spectrum shows
that the first-order transition to the Wigner crystal occurs
at $\nu\sim 1/7$.

With decreasing magnetic field, electrons occupy higher
Landau levels.
In high Landau levels, the one-particle wavefunction
extends over space, generating effective long-range
exchange interactions between the electrons. 
The long-range interaction 
prefers CDW ground states and Hartree-Fock calculations
predict various CDW states called stripe and bubble\cite{Kou}.
These CDW states are confirmed by the DMRG calculations 
as shown in figures \ref{Fig_CDW} (b) and (c).
Although the CDW structures are similar to those
obtained in the Hartree-Fock calculations, the ground-state energy and
the phase diagram is significantly different\cite{Shibata2}.
The DMRG results are consistent with recent experiments\cite{Stripe},
and the discrepancy is due to the quantum fluctuations neglected 
in the Hartree-Fock calculations.

\section{Summary}
In this topical review, we have reviewed the DMRG method and its 
applications to finite temperature and two-dimensional systems.
We first briefly reviewed the original DMRG method, which
enables us to calculate 
the ground-state wavefunctions and low-energy excitations 
of large-size systems with controlled accuracy. 
The essential idea of the DMRG is the restriction of
basis states using the eigenvectors of the
density matrix calculated from the ground-state wavefunction.
The truncation error of the wavefunction is 
estimated from the eigenvalues of the density matrix and
we can extend the size of system with controlled accuracy.

This idea of the DMRG was applied to the QTM
in one-dimensional systems, and a reliable
method for the calculation of thermodynamic quantities
at finite temperatures has been formulated.
In this way the thermodynamic quantities
are directly obtained from the maximum 
eigenvalue of the QTM, and
temperature dependence is systematically obtained by 
extending the QTM in the $\beta$-direction. 
The dynamic correlation functions 
are calculated from the eigenvectors of the QTM 
with the analytical continuation to the
real frequency axis.

We have also shown that the DMRG method is reliably
applied to two-dimensional quantum systems in 
a magnetic field.
We use the eigenstates of free electrons in the
Landau gauge as initial basis states of two-dimensional electrons,
and map the system on to an effective
one-dimensional lattice model with long-range interactions.
It has been shown that this method is successfully
applied to quantum Hall systems and
the existence of various ground states has been confirmed.

\ack
The author thanks Professor Daijiro Yoshioka for suggesting
the application of the DMRG to quantum Hall systems.
The present work is financially supported by Grant-in-Aid 
from MEXT, Japan.


\section*{References}
\begin{thebibliography}{99}
\bibitem{White}White S R 1992 \PRL {\bf 69} 2863
\bibitem{White2}White S R 1993 \PR B {\bf 48} 10345
\bibitem{Nishino}Nishino T 1995 \JPSJ {\bf 64} 3598
\bibitem{Bursill}Bursill R J, Xiang T and Gehring G A 1996 
\JPCM {\bf 8} L583 
\bibitem{Wang}Wang X and Xiang T 1997 \PR B {\bf 56} 5061
\bibitem{Shibata}Shibata N 1997 \JPSJ {\bf 66} 2221
\bibitem{FT_spin}Xiang T 1998 \PR B {\bf 58} 9142
\bibitem{FT_KLM}Shibata N, Ammon B, Troyer M, Sigrist M, and Ueda K 
\JPSJ 1998 {\bf 67} 1086

\bibitem{FT_Zigzag}Maisinger K and Schollw\"ock U, 1999 \PRL {\bf 81} 445
%flustrated spin Received 9 March 1998 \"
\bibitem{FT_tJ_Lad}Ammon B, Troyer M, Rice T M and Shibata N 
1999 \PRL {\bf 82} 3855 
%Received 9 December 1998
\bibitem{FT_KLM_d}Shibata N and Tsunetsugu H 1999 \JPSJ {\bf 68} 744
%Received December 14, 1998
\bibitem{Mutou} Mutou T, Shibata N and Ueda K 1999
\PRL {\bf 81} 4939; 1999 ibid. {\bf 82} 3727
%Received 24 June 1998
\bibitem{FT_KLM_rev}Shibata N and Ueda K 1999 \JPCM {\bf 11} R1; 1999
ibid. {\bf 11} 4289
%Received 26 August 1998
\bibitem{Naef}Naef F, Wang X, Zotos X and Linden W von der 
1999 \PR B {\bf 60} 359 
%Received 14 December 1998
\bibitem{FT_Spin_Pei}Kl\"umper A, Raupach R and Sch\"onfeld F 1999 
\PR B {\bf 59} 3612; 2000 {\it Euro. Phys. J.} B {\bf 17} 51 
%for CuGeO$_3$ Spin Peierls Received 17 September 1998
\bibitem{FT_Imp_K}Rommer S and Eggert S 1999 \PR B {\bf 59} 6301
%Received 25 September 1998
\bibitem{Shibata_KLd}Shibata N and Tsunetsugu H 1999
\JPSJ {\bf 68} 3138
%Received July 19, 1999
\bibitem{Naef2} Naef F and Wang X 2000 \PRL {\bf 84} 1320
%Received 9 July 1999 Nuclear Spin Relaxation Rates in Two-Leg Spin Ladders
\bibitem{FT_zigzag_mag}Maeshima N and Okunishi K 2000 \PR B {\bf 62} 934
%Received 3 January 2000
%Antiferromagnetic Zigzag Spin Chain in Magnetic Fields at Finite Temperatures
\bibitem{FT_SpinLAD_mag}Wang X and Yu L 2000 \PRL {\bf 84} 5399
%Magnetic field effects on two-leg Heisenberg antiferromagnetic ladders
\bibitem{ammon1}Ammon B and Imada M 2000 \PRL {\bf 85} 1056
%Spin-1 chain doped with mobile S=1/2 fermions
\bibitem{ammon2}Ammon B and Imada M 2000 \JPSJ {\bf 69} 1946
%Effect of the Orbital Level Difference in Doped Spin-1 Chains
\bibitem{ammon3}Ammon B and Imada M 2001 \JPSJ {\bf 70} 547
%Doped two orbital chains with strong Hund's rule couplings 
\bibitem{Yb4_theo} Shibata N and Ueda K 2001 \JPSJ {\bf 70} 3690
\bibitem{Yb4_exp} Iwasa K, Kohgi M, Gukasov A, Mignot J -M, Shibata N,
Ochiai A, Aoki H and Suzuki T 2002 \PR B {\bf 65} 052408 
\bibitem{Sirk3}Sirker J and Kl\"umper A 2002 Europhys. Lett. {\bf 60} 262
%t-J pathInt \"
\bibitem{Sirk1}Sirker J and Kl\"umper A 2002 \PR B {\bf 66} 245102
% t-J \"
\bibitem{Maru}Maruyama I, Shibata N and Ueda K
2002 \PR B {\bf 65} 174421 
\bibitem{Sirk2}Sirker J and Khaliullin G 2003 \PR B {\bf 67} 100408
%spin-orbital
\bibitem{White3}White S R 1996 \PRL {\bf 77} 3633
\bibitem{White4}White S R and Scalapino D J 1998 \PRL {\bf 80} 1272
\bibitem{White5}White S R and Scalapino D J 1999 \PR B {\bf 60} R753
\bibitem{White6}White S R and Scalapino D J 2000 \PR B {\bf 61} 6320
\bibitem{Shibata2}Shibata N and Yoshioka D 2001 
\PRL {\bf 86} 5755
\bibitem{Shibata4}Yoshioka D and Shibata N 2002
{\it Physica} E {\bf 12} 43
\bibitem{Xiang}Xiang T 2001 \PR B {\bf 64} 104414 
\bibitem{Shibata3}Shibata N and Yoshioka D 2003
\JPSJ {\bf 72} 664

\bibitem{Xiang2}Xiang T 1996 \PR B {\bf 53} R10445
\bibitem{Nishimoto}Nishimoto S, Jeckelmann E, Gebhard F and Noack R M 
2002 \PR B {\bf 65} 165114 

\bibitem{CTM1}Nishino T and Okunishi K 1996 \JPSJ {\bf 65} 891;
1997 \JPSJ {\bf 66} 3040 
%CTM 2D classical
%\bibitem{CTM2}Nishino T and Okunishi K 1997 \JPSJ {\bf 66} 3040 
%CTM 2D classical
\bibitem{CTM3}Nishino T and Okunishi K 1998 \JPSJ {\bf 67} 3066 
%CTM 3D
\bibitem{Kem}Kemper A, Gendiar A, Nishino T, Schadschneider A and
Zittartz J 2003 \JPA {\bf 36} 29 
%Stochastic Light-Cone CTMRG: a new DMRG 

\bibitem{Maeshima}Maeshima N, Hieida Y, Akutsu Y, Nishino T
and Okunishi K 2001 \PR E {\bf 64} 016705

\bibitem{Chun}Chung M C and Peschel I 2000 \PR B {\bf 62} 4191 
%on 2D DMRG

\bibitem{DMRG_book}  Peschel I, Wang X, Kaulke M and Hallberg K (eds) 
1998 {\em Density-Matrix Renormalization} (Heidelberg: Springer)

\bibitem{Betsuyaku}Betsuyaku H 1984 \PRL {\bf 53} 629 
\bibitem{Betsuyaku2}Betsuyaku H 1985 {\it Prog. Theor. Phys.} {\bf 73} 319 
\bibitem{Trotter}Trotter H F 1959 Proc. Amer. Math. Soc. {\bf 10} 545 
\bibitem{Suzuki1}Suzuki M 1976 Commun. Math. Phys. {\bf 51} 183
\bibitem{Suzuki2}Suzuki M 1985 \PR B {\bf 31} 2957
\bibitem{Koma}Koma T 1987 {\it Prog. Theor. Phys.} {\bf 78} 1213;
1989 ibid. {\bf 81} 783
\bibitem{Nomura}Nomura K and Yamada M 1991 \PR B {\bf 43} 8217

\bibitem{MEM1}Silver R N, Sivia D S and Gubernatis J E
1990 \PR B {\bf 41} 2380
\bibitem{MEM2}Gubernatis J E, Jarrell M and Silver R N
1991 \PR B {\bf 44} 6011
\bibitem{QHHS} Bonsall L and Maradudin A 1977 \PR B {\bf 15} 1959
\bibitem{Yoshioka}Yoshioka D 1984 \PR B {\bf 29} 6833
\bibitem{FQHEex}Du R R, Stormer H L, Tsui D C, 
Pfeiffer L N and West K W 1993 \PRL {\bf 70} 2944
\bibitem{Laugh}Laughlin R B 1983 \PRL {\bf 50} 1395
\bibitem{Kou}Koulakov A A, Fogler M M and Shklovskii B I 
1996 \PRL {\bf 76} 499
\bibitem{Stripe}Lilly M P, Cooper K B, Eisenstein J P, 
Pfeiffer L N and West K W 1999 \PRL {\bf 82} 394
\end{thebibliography}
\end{document}